\documentclass[fleqn,usenatbib]{mnras}

\usepackage{newtxtext,newtxmath}

\usepackage[T1]{fontenc}
\usepackage{ae,aecompl}

\usepackage{subfig}
\usepackage{graphicx}	
\usepackage{amsmath}	
\usepackage{siunitx}    

\usepackage{booktabs}
\usepackage{longtable}

\usepackage[dvipsnames]{xcolor}
\newcommand{\AC}{\textsc{Astrocook}}
\newcommand{\lya}{Lyman-$\alpha$\ }
\newcommand{\lyb}{Lyman-$\beta$\ }
\newcommand{\kms}{\SI{}{\km\per\second}}
\def\ms{$ \rm m~s^{-1}$}
\def\cms{$ \rm cm~s^{-1}$}
\newcommand{\mincir}{\ \raise -2.truept\hbox{\rlap{\hbox{$\sim$}}\raise5.truept
    \hbox{$<$}\ }}
\newcommand{\NHI}{${N_{\rm HI}}$}
\newcommand{\s}{{\rm s}}

\title[The small-scale structure of the IGM]{Probing the small scale structure of the Inter-Galactic Medium with ESPRESSO: spectroscopy of the lensed QSO UM673
\thanks{Based on observations collected at the European Southern Observatory, Chile, under the programme 0104.A-0468(A).}
}
\author[S. Cristiani et al.]{
\parbox[t]{\textwidth}{
Stefano Cristiani$^{1,2,3 }$\thanks{E-mail: stefano.cristiani@inaf.it},
Guido Cupani$^1$,
Andrea Trost$^{1,4}$,
Valentina D'Odorico$^{1,2}$,
Francesco Guarneri$^{1,4}$,
Gaspare Lo~Curto$^{5}$,
Massimo Meneghetti$^6$,
Paolo Di Marcantonio$^1$,
Jo\~ao P. Faria$^{10}$,
Jonay I. Gonz\'alez Hern\'andez$^{8,9}$, 
Christophe Lovis$^{7}$,
Carlos J.A.P. Martins$^{10,11}$, 
Dinko Milakovi\'c$^{2,1,3}$, 
Paolo Molaro$^{1}$,
Michael T. Murphy $^{12,2}$, 
Nelson J. Nunes$^{14}$, 
Francesco Pepe$^7$,   
Rafael Rebolo$^{8,9}$,
Nuno C. Santos$^{10,13}$,
Tobias M. Schmidt$^7$,   
S\'ergio G. Sousa$^{10}$, 
Alessandro Sozzetti$^{15}$,
Mar\'ia Rosa Zapatero Osorio$^{16}$.}
\vspace*{1pt}\\
$^{1}$ INAF--Osservatorio Astronomico di Trieste, Via G.B. Tiepolo, 11, 34143 Trieste, Italy \\
$^{2}$ IFPU--Institute for Fundamental Physics of the Universe, via Beirut 2, I-34151 Trieste, Italy \\
$^{3}$ INFN-National Institute for Nuclear Physics, via Valerio 2, I-34127 Trieste\\
$^{4}$ Dipartimento di Fisica dell'Universit\`a  di Trieste, Sezione di Astronomia, Via G.B. Tiepolo, 11, I-34143 Trieste, Italy \\
$^{5}$  European Southern Observatory, Alonso de Cordova 3107, Vitacura, Santiago, Chile \\
$^{6}$  INAF -- Osservatorio di Astrofisica e Scienza dello Spazio, Via Gobetti 93/3, 40129 Bologna, Italy \\
$^7$  D\'epartement d'Astronomie de l'Universit\'e de Geneve, Chemin Pegasi 51, 1290 Versoix, Switzerland \\
$^8$ Instituto de Astrof\'isica de Canarias (IAC), Calle V\'ia L\'actea s/n, 38205 La Laguna, Tenerife, Spain\\
$^9$ Departamento de Astrof\'isica, Universidad de La Laguna (ULL), 38206 La Laguna, Tenerife, Spain \\
$^{10}$ Instituto de Astrof\'isica e Ci\^encias do Espa\c{c}o, CAUP, Universidade do Porto, Rua das Estrelas, 4150-762 Porto, Portugal \\
$^{11}$ Centro de Astrof\'{\i}sica da Universidade do Porto, Rua das Estrelas, 4150-762 Porto, Portugal \\
$^{12}$  Centre for Astrophysics and Supercomputing, Swinburne University of Technology, Hawthorn, Victoria 3122, Australia \\
$^{13}$ Departamento de F\'isica e Astronomia, Universidade do Porto, Rua Campo Alegre, 4169-007 Porto, Portugal \\
$^{14}$ Instituto de Astrof\'isica e Ci\^encias do Espa\c{c}o, Faculdade de Ci\^encias da Universidade de Lisboa,
Campo Grande, PT1749-016 Lisboa, Portugal \\
$^{15}$ INAF -- Osservatorio Astrofisico di Torino, via Osservatorio 20, 10025 Pino Torinese, Italy \\
$^{16}$ Centro de Astrobiolog\'ia (CSIC-INTA), Carretera de Ajalvir km 4, E-28850 Torrej\'on de Ardoz, Madrid, Spain
}
\date{Accepted XXX. Received YYY; in original form ZZZ}
\pubyear{2024}
\begin{document}
\label{firstpage}
\pagerange{\pageref{firstpage}--\pageref{lastpage}}
\maketitle
\sisetup{range-phrase=\textrm{--}}
\begin{abstract}
The gravitationally lensed quasar J014516.6-094517 at $z=2.719$ has been observed with the ESPRESSO instrument at the ESO VLT to obtain high-fidelity spectra of the two images A and B with a resolving power $R=70000$.
At the redshifts under investigation ($2.1 \mincir z \mincir 2.7$), the Lyman forests along the two sightlines are separated by sub-kiloparsec physical distances and exhibit a strong correlation.
We find that the two forests are indistinguishable at the present level of signal-to-noise ratio and do not show any global velocity shift, with the cross-correlation peaking at $\Delta v = 12 \pm 48$ \ms. The distribution of the difference in velocity of individual \lya features is compatible with a null average and a mean absolute deviation of 930 \ms. Significant differences in \NHI\ column density are not detected, putting a limit to the RMS fluctuation in the baryon density on $\mincir 1$ proper kpc scales of $\Delta \rho / \rho \mincir 3$\%.
On the other hand, metal lines show significant differences both in velocity structure and in column density.
A toy model shows that the difference in velocity of the metal features between the two sightlines is compatible with the the motions of the baryonic component associated to dark matter halos of typical mass $M\simeq 2\times 10^{10} M_\odot$, also compatible with the observed incidence of the metal systems.
The present observations confirm the feasibility of the Sandage test of the cosmic redshift drift with high-fidelity spectroscopy of the Lyman forest of distant, bright quasars, but also provide an element of caution about the intrinsic noise associated to the usage of metal features for the same purpose.
\end{abstract}
%
%
%
\section{Introduction}
Over the past two decades the understanding of the
intergalactic medium (IGM), the  main baryonic component of the cosmic
web, has greatly improved. Advances have taken place both in observations (e.g. thanks to the availability of new spectroscopic facilities at VLT and Keck) and numerical simulations.
A fairly comprehensive view has been developed
of the overall properties of the baryon field, such as
dynamics, thermal evolution, radiation environment, chemical enrichment, and the manner in which they are influenced by cosmological parameters.

The fundamental issue of the physical extent of the Lyman forest absorbers has been addressed by several investigations
\citep[e.g.][]{Weymann83, Smette92, Crotts98, Dodorico98, Dodorico02, Rauch05}. These studies have generally found that the absorbers exhibit significant sizes, typically spanning few hundred kiloparsecs. 
Both observations 
\citep[e.g.][]{Smette92, Smette95} 
and simulations 
\citep[e.g.][]{Bolton17}
have shown that these absorbing structures are constituent elements of the cosmic web 
\citep{Bond96} 
and are anticipated to experience, on average, the effects of the universal Hubble expansion \citep{Dave99}.

Observations of the velocity field in the Lyman forest at $z \gtrsim 2$ give us insights into the early stages of structure formation, 
when overdense regions break away from the universal expansion and begin to collapse under the influence of gravity,
revealing how the gaseous cosmic web actually evolves, as a function of size, redshift, and density.
We may reasonably expect that the cosmic web, whose
large-scale structure can be classified as halos, filaments, and voids
\citep{Martizzi2019}, should follow the Hubble flow on
large (Mpc) scales, i.e. at least on scales larger than the typical coherence length of these structures.
At intermediate scales (of order 100 kpc) 
the influence of gravitational collapse tends to become more pronounced,
and galactic and sub-galactic potential wells may convey kinetic energy to the gas.  Conversely, on smaller scales, below the sub-kiloparsec range, the dominant sources of kinetic energy and momentum are expected to arise from stellar evolution and gasdynamical processes within the interstellar medium (ISM), specifically supernova remnants and stellar winds.
Earlier observations of small-scale structure in \lya forest
systems \citep{Rauch01a, Rauch05} have shown that there is also a trend of the
motions to increase in strength with increasing density, e.g., the
higher density gas appears to be more turbulent than the typical low-density
\lya forest gas.

Hydrodynamical simulations (see \cite{Vogelsberger2020} for a review)
offer the possibility to investigate in detail the 3-D distribution, kinematics, temperature, and chemical composition of the gas, the gravitational back-reaction onto the dark matter due to the redistribution of baryons and the feedback deriving from the formation of structures. 
However, to capture the large-scale structure, which requires box sizes with $L > 100$ Mpc, while also resolving the kiloparsec- and subkiloparsec-scale size of gas clumps is a formidable computational challenge \cite[e.g.][]{Cain2023, Katz2023}.

Along the line of sight directed toward a remote source, such as a quasar, gamma-ray burst, or galaxy, each discrete region of the IGM exerts preferential absorption of specific light wavelengths, due to the presence of the various chemical elements. The scrutiny of these absorption spectral lines enables an investigation into the spatial distribution, motions, chemical enrichment processes, and ionization histories of gaseous structures.
In particular, when a structure is pierced by two nearby lines of sight,
the absorption feature's position will be shifted between the two spectra, with the shift amplitude depending on the Hubble expansion and the line of sight separation. Additionally, shifts can be caused by peculiar motions, e.g. gravitational collapse and a wide range of other processes including galactic feedback and systematic rotation.
We would like to understand
the origin of the observed motions, focusing on the smaller
(kiloparsec and sub-kiloparsec) scales,
which have the potential to shed light on the feedback processes
governing the formation of galactic structures. 
\begin{figure*}
\includegraphics[width=0.99\textwidth]{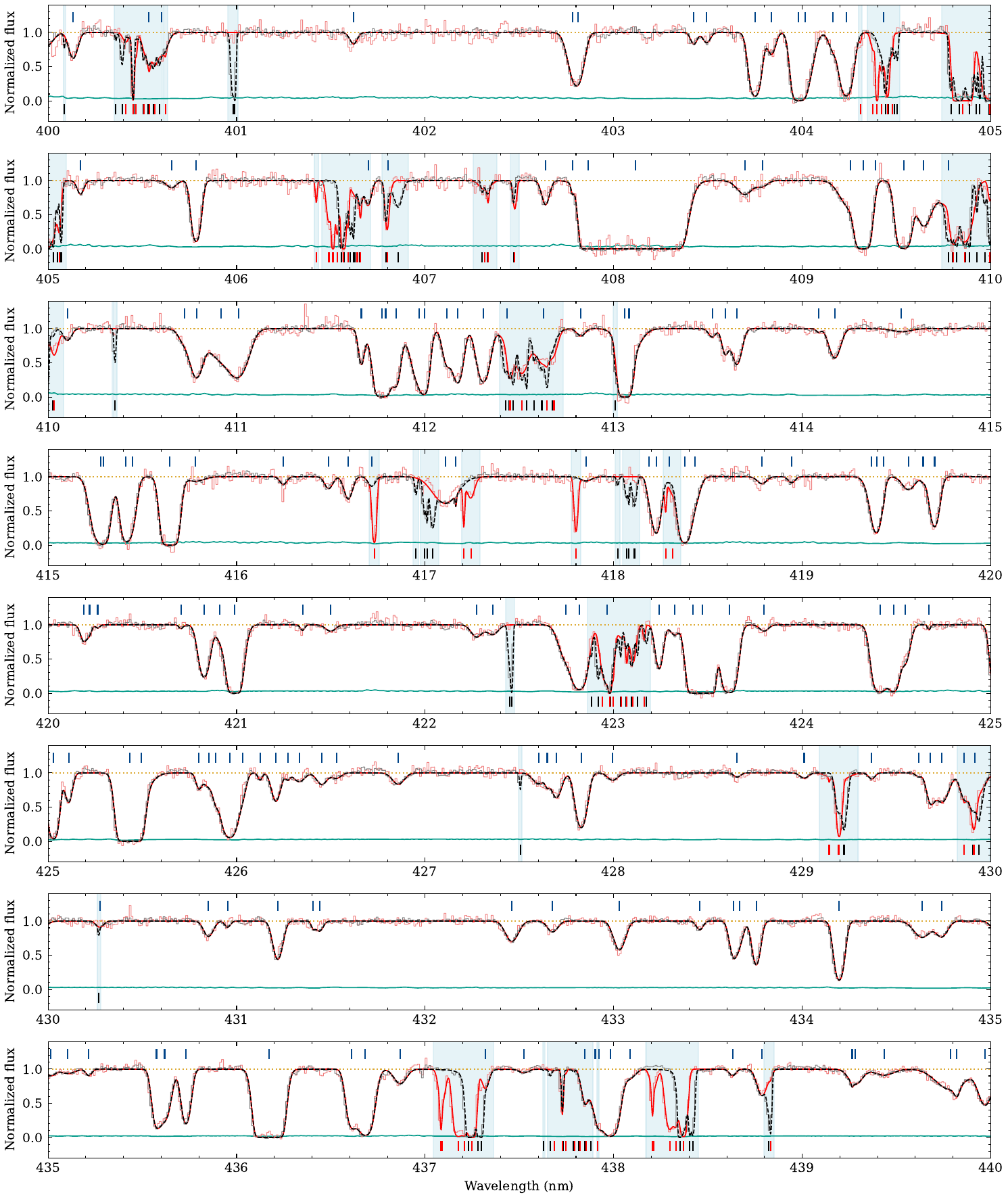}
    \caption{a) A section of the spectrum of UM 673, normalized in transmission. The spectra of the A
    (grey solid line) and B (light red  solid line) images are plotted on top of each other. The absorption models for the two images are also shown: A - black dashed line, B - red solid line. 
    The teal line shows the noise in the spectrum of the image B,
    predominant with respect to the component A. The position of the \lya components is shown with blue vertical ticks above the spectrum, while the black and red ticks below it define the positions of the metal components in image A and B, respectively. 
    The vertical blue shaded regions in both segments define areas within the \lya forest where masking has been applied due to metal contamination, (as discussed in \autoref{Sect:LineList}).
The two spectra appear indistinguishable within the noise, except for features corresponding to metal lines.}
    \label{Fig:LyaForest1}
     \ContinuedFloat
\end{figure*}

\begin{figure*}
\includegraphics[width=\textwidth]{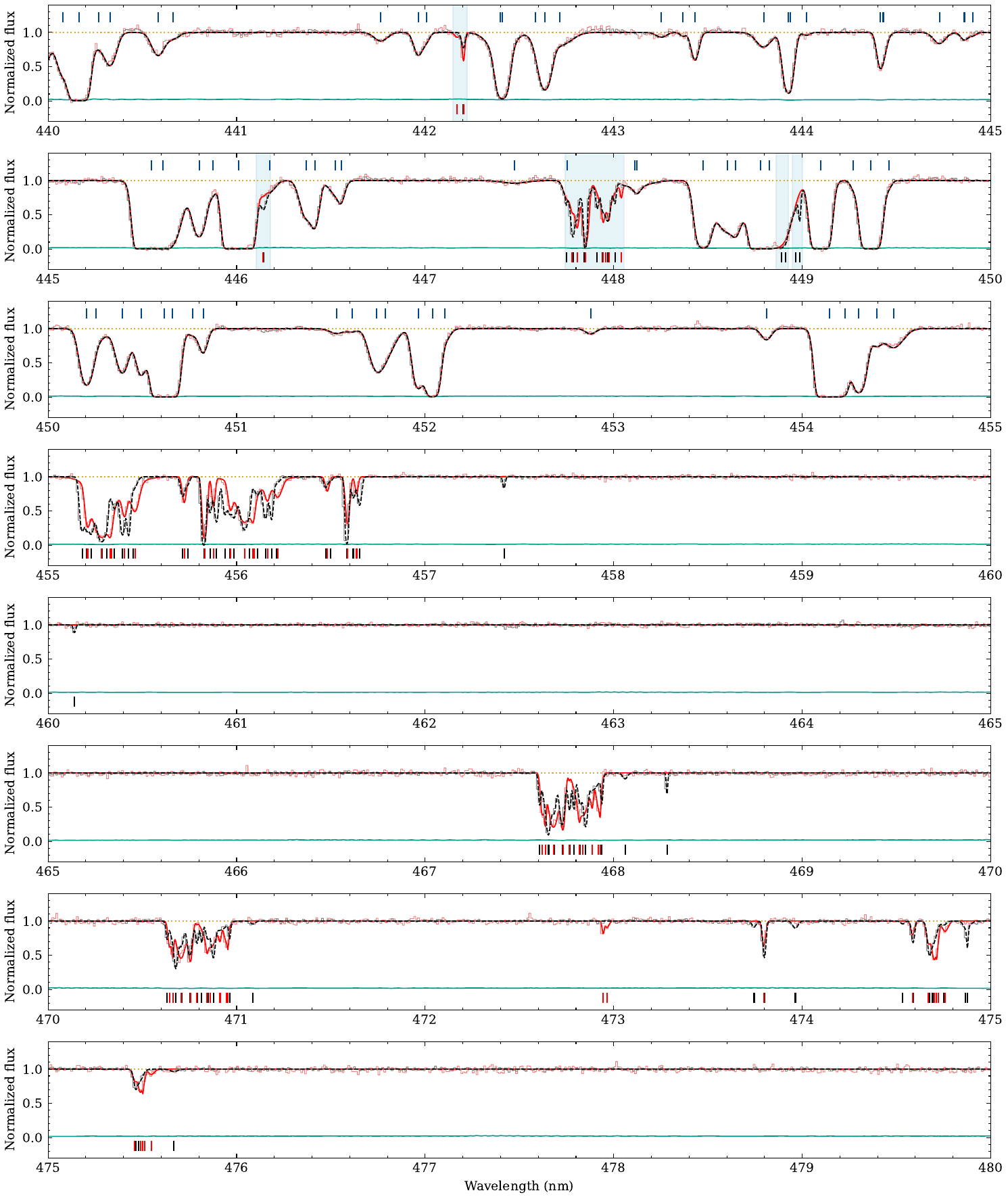}
    \caption{b) Continuation of \autoref{Fig:LyaForest1} up to $\SI{480}{\nm}$}
    \label{Fig:LyaForest2}
\end{figure*}
Another goal of the present observations is to test whether systematic
variations of the dynamical and ionization state might affect the Sandage Test of the cosmic redshift drift \citep{Sandage62} in future experiments using the Lyman forest as a probe \citep{Loeb98, Cristiani07, Liske08}.
The redshift drift ($\dot{z} = {{\rm d}z}/{{\rm d}t_{\rm obs}}$) is a small, dynamical change in the redshift of objects following the Hubble flow. Measuring it provides a direct, real-time and model-independent mapping of the expansion rate of the universe. It is fundamentally different from other cosmological probes: instead of mapping our (present-day) past light cone, it directly compares different past light cones. This measurement is a flagship objective of the Extremely Large Telescope (ELT) , specifically observing the Lyman forest of QSOs with its high-resolution spectrograph, ANDES \citep{ANDES_WP_2023}. The effect is tiny,
expected to be of order \cms yr$^{-1}$ at
the redshifts of interest, and to carry out the measurement, 
high-accuracy cosmological probes have to be observed with high-fidelity instruments. 

Throughout this paper we adopt a $\Lambda$CDM Planck 2018 cosmology 
with ${\rm H_o} = 69.6$ \kms ${\rm Mpc^{-1}}$, $\Omega_m = 0.286$ and $\Omega_{\Lambda} = 0.714$. At a typical redshift of $z=2.5$, 1 arcsec on the sky corresponds to 8.225 proper kpc \citep{Planck18}.

\section{Target Selection}
We have selected, as a target to investigate the small-scale structure of the IGM,
the brightest known lensed QSO in the Southern sky 
for which the Lyman forest is conveniently observable from the ground ($z>2.5$) and 
the separation of at least two images of the lens is above $2$" (in order to avoid
flux contamination, see \autoref{Sect:Observations}). 
J014516.6-094517 A, B is a lensed QSO with emission redshift $z=2.719$
and images separated on the sky by $2.24"$. 
The QSO, also known as UM673, was discovered by \cite{MacAlpine82}
and was classified as a double lensed QSO by \cite{Surdej87}. 
\cite{Surdej88} showed the presence of a lensing galaxy at $z = 0.49$,
as later confirmed by \cite{Eigenbrod07}.

The image A has an apparent magnitude around $R \simeq 16.2$, while the image B has $R\simeq 18.3$, with an observed variability
of about $0.1$ mag \citep{Koptelova10}.
The extinction-corrected flux ratio between components A and B has been estimated to be about 2.14 mag \citep{Koptelova14}. 
The Lyman-a forests along the lines of sight towards UM673A,B cover the redshift range from 2.2 to 2.7 and correspond, given the geometry of the lens (see \autoref{Sect:LensModel}), to transverse scales from $\sim 1 h^{-1}_{69.6}$ 
physical kiloparsecs down to the few hundred pc range.

Previous measurements at intermediate resolution are available for a similar
object, RXJ0911.4+0551 \citep{Rauch05}, and show
a remarkable similarity in the Lyman forests at these scales. 
However, the available data on RXJ0911.4+0551 are
limited by the precision and stability of the ESI spectrograph at Keck, which implies that
differences in velocity between the pairs of absorption
features are fully compatible with the measurement error of about \SI{5}{\km\per\second}.
\section{Observations of UM 673 A, B}
\label{Sect:Observations}
We have taken advantage of the outstanding stability of the ESPRESSO spectrograph \citep{ESPRESSO2021} to significantly improve the precision of the measurements with respect to previous studies of the small-scale structure of the IGM \citep[e.g.][]{Rauch01b,Rauch05}, i.e. to reach a
sensitivity of about \SI{0.5}{\km\per\second} per spectral feature (assumed to be a
\lya line with a Doppler parameter of $\sim \SI{20}{\km\per\second}$).

Observations were carried out on Nov 26, 2019 at ESO Paranal, in 4UT mode, i.e. incoherently combining the light of the 4 unit telescopes (UTs) of the ESO Very Large Telescope (VLT). 
ESPRESSO is a fiber-fed spectrograph
with the fiber size corresponding to 1" on the sky.
Three exposures of 3600s were obtained on the image B and two of 1200s on image A. 
The different exposure times were chosen in order to (partially) compensate the ($\sim 9$) flux ratio between the two images, as measured in the $g$ band by the DES DR2 Survey 
\citep[][$g_A=16.8965$ and $g_B=19.2846$]{DES_DR2}, the rough expectation being of a $\sim \sqrt{2}$ ratio of the final SNR between the image A and B combined spectra (see Sect. \ref{Sect:Reduction}).

The seeing conditions were on average around 0.7" and
the expected fiber flux contamination in the observation of the image B from the image A turns out to be negligible (less than $10^{-7}$, assuming a 2-D gaussian for the seeing). The CCD detectors were binned by a factor 
$8 \times 4$ along the X and Y (dispersion) directions, respectively. The resolving power of ESPRESSO in the 4UT mode is $R=70000$, independent of the wavelength, as measured by \cite{ESPRESSO2021}.
We modeled the instrument LSF with a Gaussian profile with FWHM corresponding to a resolution of 70K. Though necessarily an approximation, the profile provides within the scope of this paper a fully sufficient description of the LSF of the unresolved ThAr lines we extracted from the calibration lamps.

Data of UM 673 A, B are also found in the KODIAQ database \citep{KODIAQ2}, obtained with the HIRES spectrograph at Keck, for a total of 14400 and 28600 s
for the image A and B, respectively. More details about these 2005 and 2008 observations are given in \cite{Cooke2010}.

\section{Data Reduction and Analysis}
\label{Sect:Reduction}
The observed spectra were reduced with the ESPRESSO Data Reduction Software\footnote{\url{https://www.eso.org/sci/software/pipelines/}, version 3.0.0.} \citep[DRS, ][]{DiMarcantonio2018}. 
Data reduction included wavelength calibration (converted to the Solar System barycentric frame in vacuum), sky and background subtraction, and optimal extraction of the echelle orders along the cross-dispersion direction (i.e. without any rebinning in the dispersion direction). 

The reduced exposures were merged into two spectra, one for image A and one for image B (\autoref{Fig:LyaForest1}), with the ESPRESSO Data Analysis Software\footnotemark[\value{footnote}] \citep{Cupani19} and with the \AC{} package\footnote{\url{https://github.com/DAS-OATs/astrocook}} \citep{Cupani20}. The exposures were first adjusted by the ratio of their median flux densities to account for discrepancy in their relative photon counts; the discrepancy was in all cases within 15\% and showed no significant dependence on wavelength. The exposures were then combined by defining a final wavelength grid and computing a weighted average of the contributions to each bin of this grid from the original pixels of the echelle orders. The grid was chosen to be identical for both images, to allow pixel-by-pixel comparison of the two sightlines, with a wavelength range  $\SIrange{380}{780}{nm}$ and a log-wavelength step corresponding to $\delta v = \SI{2.0}{\km\per\second}$ ($\delta\lambda\simeq\SI{3.7}{10^{-3} nm}\times\lambda/\SI{550}{nm}$), matching the typical size of the detector pixels for the adopted instrumental setup. 

\begin{figure}
\includegraphics[width=0.5\textwidth]{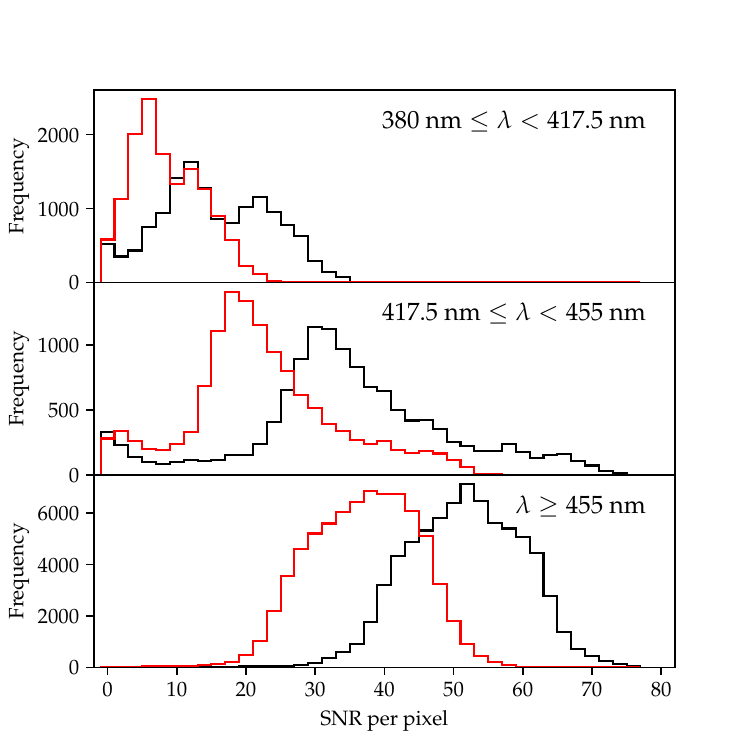}
    \caption{SNR per 2 \kms\ bin in different spectral ranges, for image A and image B (black and red, respectively). The top and centre panel cover the two halves of the \lya forest, while the bottom panel covers the region redwards from it.}
    \label{Fig:snr}
\end{figure}

The median signal-to-noise ratio (SNR) at continuum for image A and image B is respectively $\simeq 70$ and $\simeq 50$ per 4.28 \kms\ resolution element ($\simeq 50$ and $\simeq 35$ per spectral bin, with $\simeq 2$ bins per resolution element), ranging between $\simeq 10$ and $\simeq 80$ per spectral bin in the Lyman-$\alpha$ forest and rapidly decreasing bluewards of $\SI{400}{nm}$ in the observed frame. This values are consistent with the expectations, given the observing conditions, as confirmed by the ESPRESSO Exposure Time Calculator\footnote{\url{https://www.eso.org/observing/etc/bin/gen/form?INS.NAME=ESPRESSO+INS.MODE=spectro}}. For comparison, the HIRES spectra were binned with a log-wavelength step $\delta v = \SI{2.5}{\km\per\second}$ and achieved a median SNR at continuum of $\simeq$$130$ and $\simeq$$60$ ($\simeq$$75$ and $\simeq$$35$ per spectral bin, given $\simeq$$3$ bins per resolution element) for image A and image B respectively. A histogram of the SNR per bin in different spectral ranges is given in \autoref{Fig:snr}.

The emission continuum in the ESPRESSO spectra was modeled by \AC{} recipes computing a running average of the flux within a $\simeq\SI{400}{\km\per\second}$ window and rejecting outliers at $3\sigma_\mathrm{clip}$, with $\sigma_\mathrm{clip}$ the formal error on flux. Only outliers below the running average were rejected, and the procedure was iterated until convergence. The continuum model was obtained from the final running average, smoothed by a gaussian kernel with $\sigma_\mathrm{smooth}=\SI{300}{\km\per\second}$. 
In this way the observed spectra have been normalized to the respective continua and represent the transmission of the IGM
(see Figure~\ref{Fig:LyaForest1}).

\subsection{Modeling of the absorption features}
\label{Sect:LineList}

We have used \AC{} to model all the absorption features with composite Voigt profiles. The modeling involves three main steps:
\begin{enumerate}
    \item \emph{Detection and identification of the absorbing species.} The standard approach of {\textbf \AC{}} combines detection and identification of the absorbing species in a single procedure. To this purpose, a reference list of ionic transitions typical of the IGM is used. If there are $n$ transitions in the list, $n$ realizations of the spectrum are obtained by converting the spectral wavelengths into redshifts: $z+1 = \lambda/\lambda_{\textrm{r},i}$, where $\lambda_{\textrm{r},i}$ is the rest wavelength of each transition, for $1\leq i \leq n$. The normalized flux $f_\lambda$ is converted into a measure of {\it prominence} for the spectral features, which expresses how much a feature stands out from the local continuum in terms of the local noise: $p_\lambda\propto(1-f_\lambda)/\sigma_\lambda$, where $\sigma_\lambda$ is the error on normalized flux. The prominence profiles of the $n$ realizations are then multiplied in redshift space: in this way, prominence spikes that show a redshift coincidence across the list of transitions are reinforced, confirming the detection and providing its identification as a candidate absorption system.
    In the specific case of the UM673, the automatic procedure was facilitated by pre-selecting the absorption-affected regions and by visually inspecting the cases of multiple identification.
    \item \emph{Definition of the model components.}
    Each systems is modeled with a variable number of components. The initial number is the number of local maxima in the prominence profile. After a first fit, further components are iteratively added to adjust the model, until no significant residuals are left  (see below).
    \item \emph{Fitting of the model}: a Voigt profile is defined for each component. Each profile is parameterised by an atomic species, with corresponding atomic parameters, and three free parameters. Atomic parameters (such as laboratory wavelength and oscillator strength) are provided within the \AC{} package, and are a compilation of several literature sources. The free parameters are the redshift of the absorber, $z$, its column density, generally expressed as $\log(N)$\footnote{To simplify the notation, throughout the paper we use $\log{(N)}$ in place of $\log_{10}{(N/{\rm cm^{-2}}})$}, and a Doppler parameter describing the line broadening \textit{b} (\kms{})
    accounting for thermal broadening. \AC{} also requires information about the instrumental profile, which is convolved with the intrinsic line profile in the observed spectrum. Here it is assumed to be a Gaussian, with a full width at half maximum of {4.28} \kms, corresponding to the nominal resolution of the adopted ESPRESSO configuration. \AC{} optimises the free parameters using nonlinear least squares minimisation. It then reports the best-fit value for the free parameters, the corresponding errors, the $\chi^2$ and the reduced $\chi_\nu^2$ (i.e., the $\chi^2$ divided by the number of degrees of freedom). While building the model, changes were made only based on $\chi^2_\nu$ (that is, a component was added or removed only if it was deemed necessary to lower the $\chi_\nu^2$ of the fit).
    A composite Voigt profile is then created by combining the overlapping profiles of all components and fitted to the spectrum. 
\end{enumerate}

In practice, we started from the secure identifications of metal doublets in the region redwards of the \lya emission peak, and looked for associated absorption both outside and within the Lyman forest. We then masked all the identified metal absorbers and looked for differences in the Lyman forest absorption pattern along the two sightlines. Under the assumption that \lya and \lyb could not be too dissimilar in the two spectra, we looked for significant spikes in the difference spectrum (i.e. contiguous regions of several wavelength bins where the absolute value of the difference was larger than 5 times the local error) and added them to the mask of metal absorbers, even when a secure identification was not possible. \lya and \lyb lines were first modeled on a combination of the two masked spectra and then re-fitted to the individual sightlines. The procedure was iterated several times, adding new identifications at each iteration, until a solid assessment of the large majority of features was achieved.

A total of 332 \lya/$\beta$ components were modeled on both spectra, while we identified 371 metal components on A and 232 metal components on B.
A list of the components with their best-fitting parameters is given in Table \ref{lya_list}, Table \ref{metal_list_A}, and Table \ref{metal_list_B}. 
\begin{figure}
 \includegraphics[width=240pt]{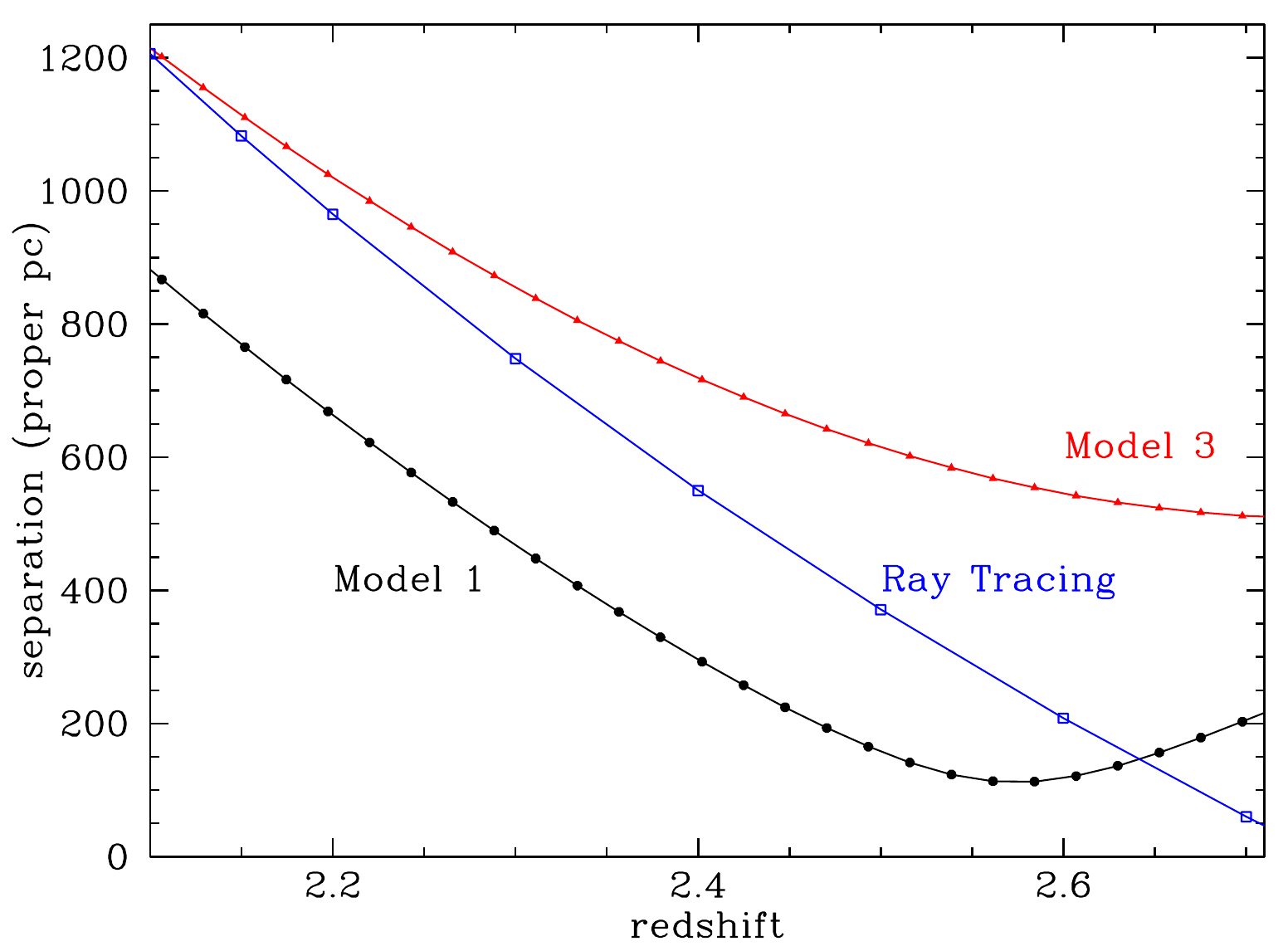}
\caption{Separation of the lines of sight to the images A and B as a function of the redshift, computed according to the Model 1 and 3 of \citet{Koptelova14}. The blue line is derived from a simple ray-trace equation \citep{Smette92} assuming for the redshift of the QSO  $z_S = 2.7434$ \citep{Cooke2010}, for the redshift of the lens $z_L = 0.493$ and a separation $\theta = 2.22$ arcsec.}
    \label{Fig:Separation}
\end{figure}
\section{The Lens Model}
\label{Sect:LensModel}
Determining the physical separation between the lines of sight associated with images A and B requires a model of the gravitational lens.
In this regard, we have employed the methodology presented by \cite{Koptelova14}, which incorporates constraints derived from the positional information of the two quasar components, their flux ratio, and a time delay of $89 \pm 11$ days, modeling the lensing galaxy as a singular isothermal ellipsoid.
Nevertheless, further possibilities exist regarding the presence of additional factors, 
such as an external shear or a shear at the location of one of the galaxies observed with the Hubble Space Telescope (HST) in close proximity to the line of sight (see \cite{Koptelova14} for more detailed information). While these specific details are not crucial for the analysis carried out in this paper, we have adopted the average value between Model 1 and Model 3\footnote{Model 3 presented by Koptelova et al., 2014, uses seven parameters and is therefore under-constrained.} as the physical separation between the lines of sight associated with images A and B. The uncertainty is estimated as the half-difference between the two aforementioned models.
\autoref{Fig:Separation} shows the range of separations encompassed within the redshift range relevant to the \lya forest ($2.2 <z< 2.7$) where the average separation spans approximately $<S> \sim 500~  h^{-1}_{69.6}$ 
physical parsecs.
\begin{figure*}
\includegraphics[width=250pt]{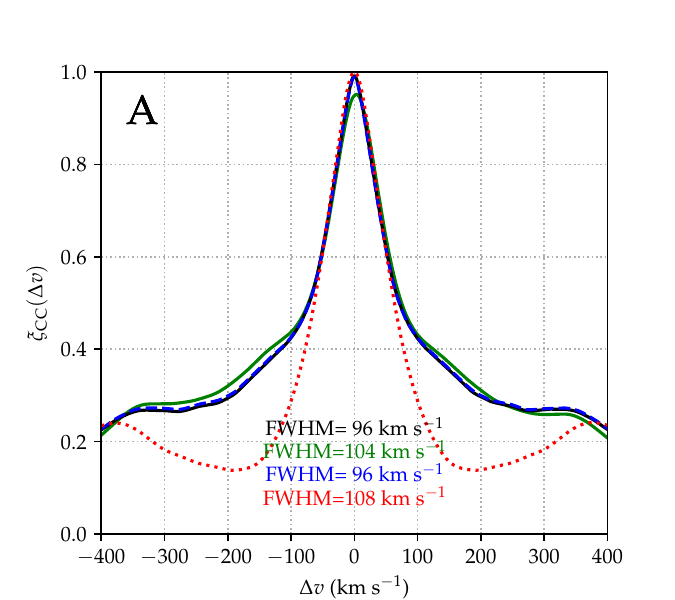}
\includegraphics[width=250pt]{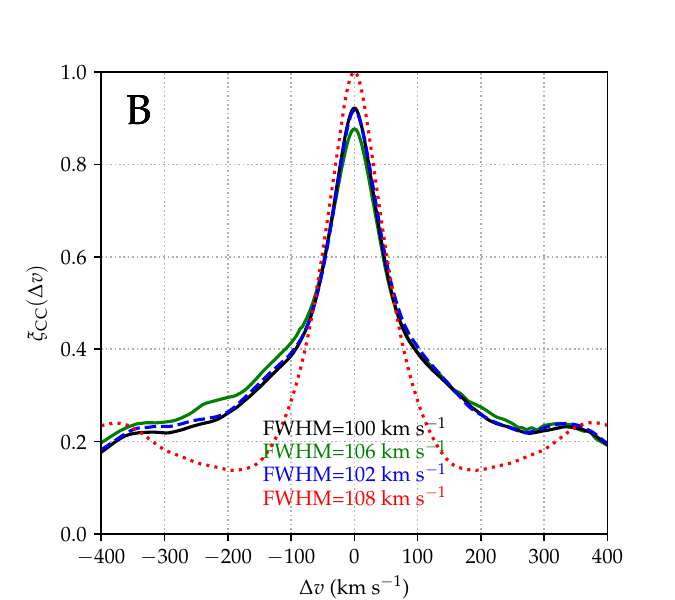}
\includegraphics[width=250pt]{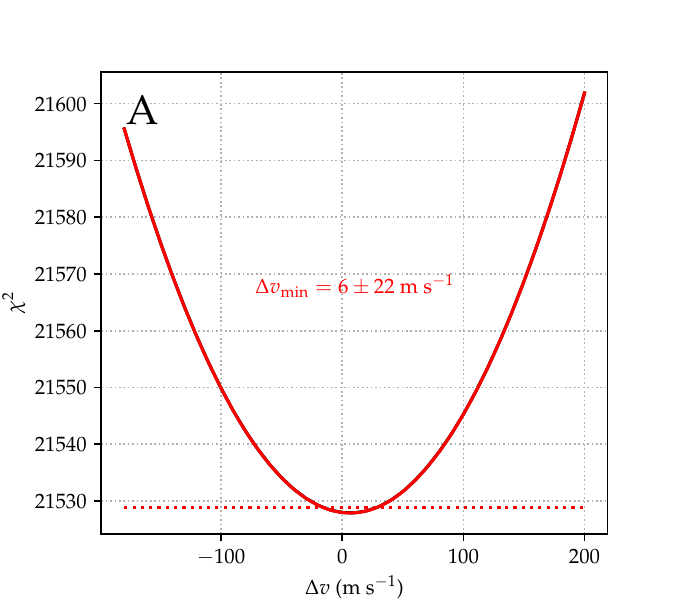}
\includegraphics[width=250pt]{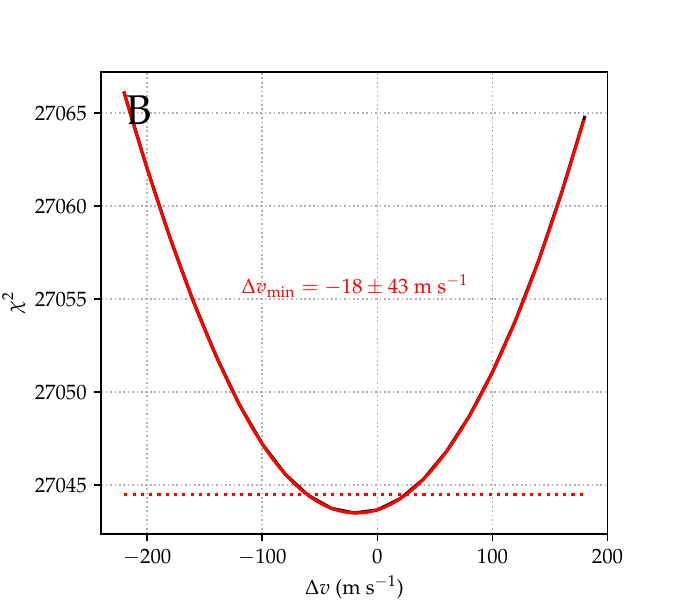}
    \caption{Cross-correlation function (CCF) $\xi_\mathrm{CC} (\Delta v)$,
    between data and model in the \lya forest ($\SIrange{380}{454}{\nm}$ observed) for the two sightlines to UM 673. Top panels: CCF profiles for the ESPRESSO observations (solid black line), the HIRES observations (solid green line), mock spectra with the same SNR of the ESPRESSO observations (dashed blue line), and mock spectra with random systems correlated with themselves at infinite SNR (dotted red line). The FWHM of the three distributions are printed with matching colors. Bottom panels: $\chi^2$ profiles around the CCF maxima. The red line (2-degree polynomial fit) almost perfectly overlaps the black line (sampled values). The shift between data model with its uncertainty, computed from the 1-$\sigma$ region around the $\chi^2$ minimum, is printed in red. The letter in the top left corner of each plot identifies one of the two sightlines.
    }
    \label{Fig:CCF}
\end{figure*}
\section {The small-scale structure of the IGM traced by the Lyman forest}
\label{Sect:SSS_Lya}
\autoref{Fig:LyaForest1}  shows the transmission of the \lya forest in a section of the spectra of UM 673 A and B.
Notably, the spectra exhibit minimal discernible differences, facilitating a straightforward correspondence between the identified absorption features in both images. In the subsequent analysis, we will examine the similarity of the two lines of sight by conducting a global correlation analysis and directly comparing individual absorption features.
\subsection{Global Correlation}\label{sect:global}
A first statistical estimator of the differences/similarities in the two lines of sight is the cross-correlation function (CCF) between the observed transmission and its model, $\xi_\mathrm{CC}$, computed over the 
global region between the \lya and the \lyb in emission:
\begin{equation}
\xi_\mathrm{CC}(\Delta v)\equiv\frac{\langle(T(v)-\overline{T})\cdot(T_M(v+\Delta v)-\overline{T}_M)\rangle}{\sqrt{\langle (T(v)-\overline{T})^2\rangle\cdot\langle(T_M(v + \Delta v)-\overline{T}_M)^2\rangle}} 
\label{Eq:ccf} 
\end{equation}
where $T$ is the observed transmission along either A or B, $T_M$ is the corresponding modeled transmission, $\Delta v$ is the velocity lag, and the overline denotes average along the whole wavelength range.
The function is defined so as to satisfy $\xi_\mathrm{CC} (0)=1$.

Given the similarity between the two forests, we produced a single {\it reference model} of the \lya and \lyb lines (obtained by fitting the spectrum of sightline A) and used \AC{} to compute the CCF along both sightlines in the observed wavelength range $\SIrange{380}{455}{\nm}$ (including only the \lya forest). 
We masked the regions affected by metal contamination, which were determined with the procedure described in Section~\ref{Sect:LineList} (examples of such regions are shown as shaded areas in \autoref{Fig:LyaForest1}). 
The resulting profile of $\xi_\mathrm{CC}(\Delta v)$ are shown with solid black lines in the top panels of \autoref{Fig:CCF}. 
The CCF peaks at a $\Delta v$ very close to 0, which would indicate a null lag between the sightline and the model.
To estimate the uncertainty of this measurement, we shifted the model around the best fit position (ranging between $-200$ and $+200$ \ms\ , with a step of 20 \ms) and computed the $\chi^2$ at all positions, after resampling the model on the wavelength grid of the data. We then modeled the resulting profile with a 2nd-degree polynomial around the $\chi^2$ minimum.
The 1-$\sigma$ region around this minimum is defined by the portion of the profile that lies below the minimum $\chi^2$ plus one.
With this definition, we obtain a model-data shift of 
$\Delta v_\mathrm{min}=-6\pm {22} $ \ms for A and 
$\Delta v_\mathrm{min}=-18\pm {43} $ \ms for B.
Both values are compatible with zero 
and with each other.
Combining the two values a velocity difference of $\Delta v = 12 \pm 48$ \ms\ is measured, indicating an absence of a detectable lag between A and B.

The height of the CCF peak is lower than 1, due to the presence of noise in the spectra. Accounting quantitatively for the effect of noise in \autoref{Eq:ccf} is difficult, because of the the presence of random and systematic noise. Therefore, to assess the effect of noise on the measured CCF we resorted to simulated spectra. The ``mocks'' were constructed from the \AC{} model of the \lya absorption lines in the non-masked regions (see Section~\ref{Sect:LineList}) and  artificially degraded to match the SNR observed along the A and B sightlines, respectively, in order to account for random errors.
We took into account possible systematic errors in the continuum estimation by allowing for a relative tilt in the continuum level of the two mocks. The profile of $\xi_\mathrm{CC}(\Delta v)$ obtained from the mocks is shown with a dashed blue line in \autoref{Fig:CCF}, adopting 
a relative continuum adjustment ranging from 0 at the red end to $\sim$20 percent at the noisier blue end. Other choices of the parameters used to construct the mocks give similar results, showing that the combined effect of observational noise and modeling uncertainties are enough to explain the decrement in the CCF peak value. 

The observed FWHM of $\xi_\mathrm{CC}(\Delta v)$ is about $\SI{110}{\km\per\second}$ for both the observed spectra and the mocks. The broadening of the CCF is a consequence of the typical width of the absorption features in the \lya forest in the velocity space (with the stronger, saturated lines that tend to widen the shape of the CCF).
To assess its value, we created a synthetic sightline at (virtually) infinite SNR and populated it with 100 absorption systems. Redshifts, column densities, and Doppler broadening of the systems were chosen randomly in the ranges $2.126<z<2.734$, $13<\log(N_\textrm{H\textsc{I}}/\SI{}{\cm\squared})<14$, $10<b/\SI{}{\km\per\second}<100$. The $\xi_\mathrm{CC}(\Delta v)$ profile obtained by correlating the synthetic sightline with itself is shown with a dotted red line in \autoref{Fig:CCF}. The distribution peaks at 1, as expected in absence of noise, and its FWHM is $\SI{112}{\km\per\second}$, very close to the one of the observed distribution. 
It is noteworthy that the synthetic sightline does not account for clustering of \lya lines, which is known to have typical scales of
$100 - 200$ \kms \citep{Cristiani1997, Saitta2008, Maitra2022},
thus its CCF does not show the wider wings that are evident in the observed profiles. 

The same procedure has been applied to the HIRES spectra and is shown in Figure \ref{Fig:CCF} as a solid green line. The FWHM of this profile is comparable with the one obtained with ESPRESSO, confirming that the width of the CCF is only defined by the typical width of the \lya lines. However, the position of the peak of the HIRES CCF shows a shift of $\SI{2.9}{\km\per\second}$ between A and B (mostly due to a positive offset of A's peak with respect to 0), suggesting the presence of possible systematic effects in the wavelength calibration of the HIRES spectra.
\begin{figure*}
\includegraphics[width=\textwidth]{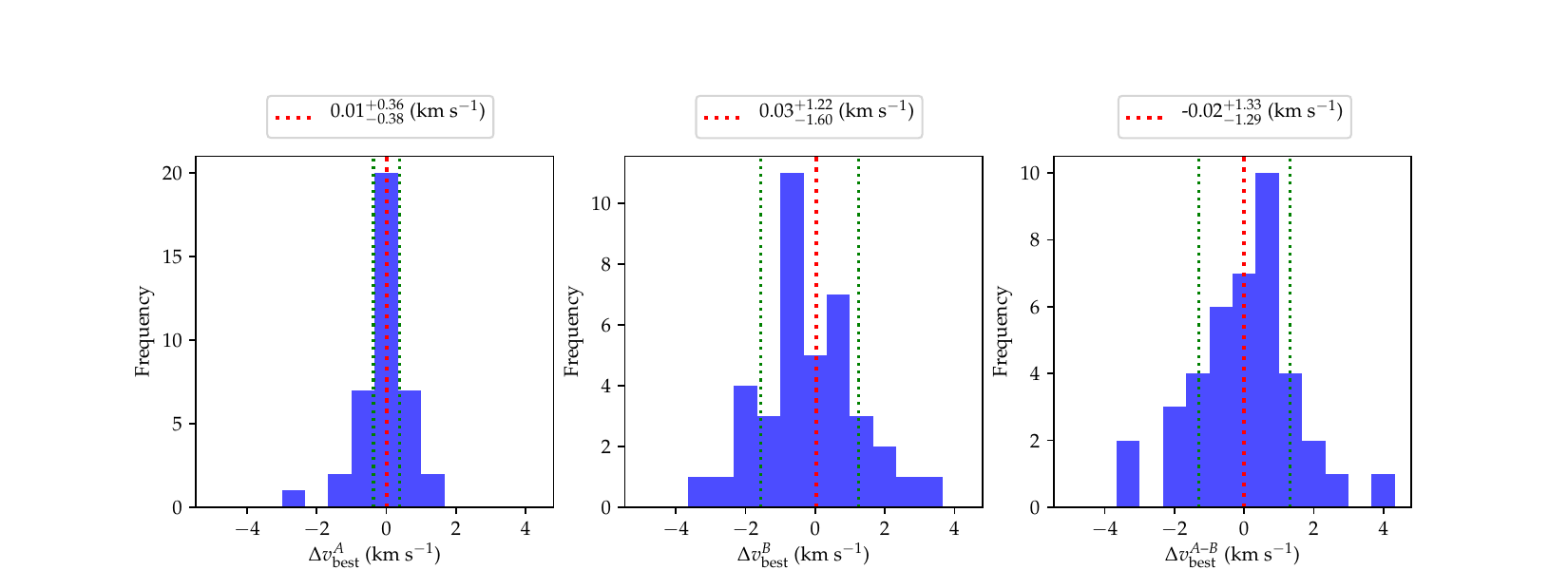}
    \caption{Distribution of the velocity shifts that maximize the cross correlation between model and data, for selected \lya features (see text). Left: sightline A; center: sightline B; right: difference between the two sightlines. The vertical dotted bars show the average and the 16-84 percentile region, corresponding to the values and confidence interval in the boxes.}
    \label{Fig:deltav}
\end{figure*}
\subsection{Comparison between individual Lyman-$\alpha$ features}
\label{sect:lyafeat}
Given the similarity between the Lyman forests of image A and B (see \autoref{Fig:LyaForest1}), it is straightforward to carry out an unambiguous, feature-by-feature comparison of the absorbers along the two lines of sight.
To this end, we used the reference model described in \autoref{sect:global}.
The {\it absorption features} to be compared between the image A and B have then been defined as the regions enclosed between two adjacent maxima in this fitted model.
This specific definition was chosen to avoid potential ambiguities linked to the decomposition of features into Voigt components, and captures the characteristics that are commonly recognized as a "feature" through visual examination.
These features represent the distinct signatures of individual absorbers, which can exhibit a simple structure, such as a single Voigt profile, or frequently consist of multiple overlapping components. By employing this definition, a total of 270 absorption features associated with \lya are identified.
Among these, in order to avoid the effects
of contamination from adjacent absorbers and low SNR,
only the 40 \emph{most reliable features} -- i.e. those with well defined boundaries (maxima $>0.85$ the continuum level) and large enough equivalent width ($>\SI{0.03}{nm}$) -- were used in the following analysis. 

\subsubsection{Differences in velocity}\label{sect:diff_in_vel}
For each feature we compared the model with the normalized flux in spectrum A and B, sliding the model in velocity space to find the velocity shifts $\Delta v_\textrm{best}^A$ and $\Delta v_\textrm{best}^B$ that gave the best correlation between the model and the data. 
The resulting distribution of $\Delta v_\textrm{best}^A$ and $\Delta v_\textrm{best}^B$, together with the distribution of $\Delta v_\textrm{best}^{A-B}=\Delta v_\textrm{best}^A-\Delta v_\textrm{best}^B$, are shown in \autoref{Fig:deltav}.
The distribution of the differences in velocity between the absorption features is compatible with a null average and a mean absolute deviation of \SI{0.93}{\km\per\second}.

If we model the \lya absorbers as slabs following the universal expansion, inclined at an angle $\theta$ with respect to the line of sight, the difference of radial velocity, $\Delta v$, between two lines of sight with a separation $l$ is expected to be:
\begin{equation}
    \Delta v = H(z) \cdot l / \tan(\theta)
\end{equation}
with an expected median $\Delta v \sim\SI{120}{\m\per\second}$, over a random distribution of $\theta$, and a typical separation of \SI{0.5}{kpc} at a $\langle z\rangle \sim 2.45$. Clearly, such a signal, if present, is still below our measurement error.
\subsubsection{Differences in column density}
\label{sec:DiffColDen}
Under the assumption that the Jeans length is a good estimate of the typical scale of the region where the density is of the order of the maximum density,  along any sightline through an absorbing cloud,
it is possible to express the observed $N_\mathrm{HI}$ column density at a given redshift as a function of the local overdensity, $\delta$ \citep{Schaye2001}:
\begin{align}
N_\mathrm{HI} \sim & ~\SI{2.7e13}{\per\centi\metre\squared} ~ (1+\delta)^{1.5-0.26 \alpha} ~
T_{0}^{-0.26} ~ \Gamma_{12}^{-1} \times \nonumber\\
& \times \left (\frac{1+z}{4}\right )^{9/2}
\left (\frac{\Omega_b h^2}{0.02}\right )^{3/2}
\left (\frac{f_g}{0.16} \right )^{1/2}
\end{align}
where 
$\delta \equiv (n_H - \bar{n}_H)/\bar{n}_H$,
$T_0$ is the temperature of the IGM at the mean density,
$\alpha$ is the index of the so-called equation of state,  $T = T_0 (1+\delta)^{\alpha}$ \citep{Hui1997},
$\Gamma \equiv \Gamma_{12} \times
10^{-12}~\s^{-1}$ is the hydrogen  
photoionization rate \citep{Bolton07},
and $f_g$ is the fraction of the mass in gas.
In this way we obtain $N_\mathrm{HI} \propto \rho_b^{\beta}$, with $\beta \equiv 1.5-0.26 ~\alpha $ and we expect to have \citep{Rauch01b}:
\begin{equation}
    \left\langle (\Delta \log \rho_b)^2\right\rangle \simeq \beta^{-2} ~ \left\langle (\log N_\mathrm{HI}^A - \log N_\mathrm{HI}^B)^2 \right\rangle
\end{equation}
In the literature various values for $\alpha$ have been reported: 1.2 \citep{Gaikwad2021}, 1.5 \citep{Telikova2019}, 1.7 \citep{Walther2019}. 
Therefore, we make the assumption that the parameter $\beta$ falls within the range of values between 1.06 and 1.19.
\begin{figure}
\includegraphics[width=8cm]{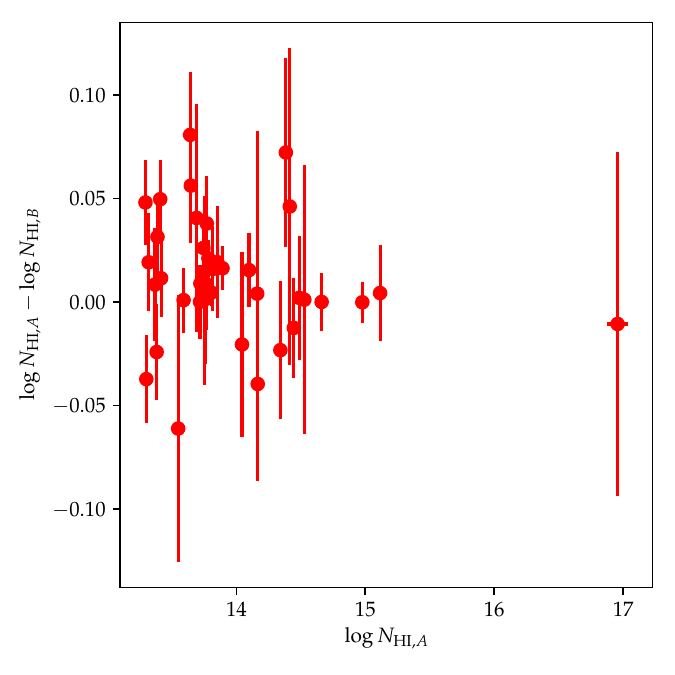}
    \caption{Differences in the logarithmic column density for pairs of \lya absorption features, plotted as a function of the model logarithmic column density.}
    \label{Fig:deltaNH}
\end{figure}
\begin{figure*}
\includegraphics[width=\textwidth]{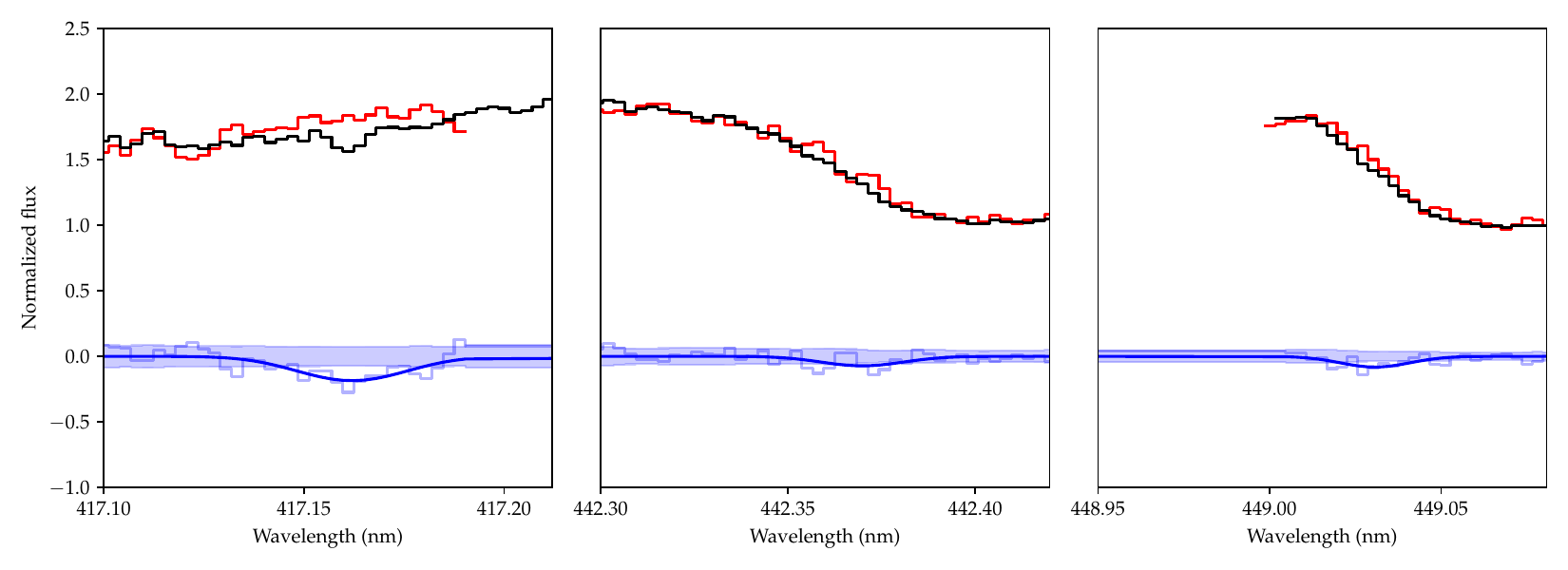}
    \caption{Features detected in the difference spectra. The normalized spectra of sightlines A and B are shown at the top (black and red lines, respectively), vertically offset by 1 unit for displaying purposes. The difference spectra $\Delta f_{A-B}$ are shown in light blue.} The corresponding best-fitting Voigt profiles are also shown with their equivalent width (dark blue). The shaded areas correspond to the 2-$\sigma$ regions around zero.
    \label{Fig:diff_feats}
\end{figure*}

In order to measure the differences in the total column densities of the 40 most reliable \lya features (listed in Table \ref{Tab:Lyfeatures}), we re-fitted their total column density, $\log N_\textrm{HI,tot}$, independently in the two lines of sight\footnote{To this end, for 
one of the H\textsc{i} components of each feature in each sightline we defined the column density as the difference between the total column density of the feature and the sum of the column densities of the other components. As a result, $\log N_\textrm{HI,tot}$ was fitted as one parameter of the composite Voigt model of the feature. The resulting value and its uncertainty are largely insensitive on the choice of the components used to model the system and provide a robust estimate.}.
The differences in logarithmic column density after re-fitting are displayed in \autoref{Fig:deltaNH} as a function of $\log N_\mathrm{HI}^A$. The distribution does not show a significant deviation from zero 
nor a significant dependence on column density. 
We obtained  $\left\langle (\log N_\mathrm{HI}^A - \log N_\mathrm{HI}^B)^2 \right\rangle = 9.4 \times 10^{-4}$.
The mean variance of $\log N_\textrm{HI}^A$ is $3.0 \times 10^{-4}$
and that of $\log N_\textrm{HI}^B$ is $1.0 \times 10^{-3}$ 
(see Table ~\ref{Tab:Lyfeatures}).
Taking into account the variances of the $ N_\mathrm{HI}$ measurements, 
the $ \left\langle (\Delta \log \rho_b)^2 \right\rangle$
is compatible with a zero value.
In order to put an upper limit to the fluctuations in baryon density, $\Delta \rho_b / \rho_b$, we consider that the standard deviation of a variance, for a normally distributed quantity, is expected to be $ \sigma_{\rm var} = \sigma^2 \sqrt{{2}/(n-1)}$, where $n$ is the number of measurements.
Then the upper limit to the typical logarithmic change in density can be computed as
$ \sqrt{\left\langle (\Delta \log \rho_b)^2\right\rangle } 
\leq \sqrt{\sigma_{\rm var}} ~\beta^{-1} =
 1.4~ {\rm or} ~1.2 \times 10^{-2}$
(depending on the assumed value for $\alpha$, 1.7 or 1.2, respectively)
on a scale of $0.2 - 1$ proper kpc; i.e., the RMS fluctuation in the baryon density, $\Delta \rho_b / \rho_b$, is less than $3.1-2.8\%$.

\subsubsection{Differences in normalized flux}
The Lyman forests observed in the images UM673A and UM673B present an opportunity for investigating the extent of the universe influenced by feedback phenomena, such as galactic winds or supernova explosions. 
As discussed in Section~\ref{sec:DiffColDen}, an analysis of the fitted Lyman features in the spectra of both images reveals no significant discrepancies. However, in order to identify subtle variations in Lyman absorption that may have eluded detection during the model fitting process, we conducted a pixel-to-pixel comparison of the normalized flux along the two lines of sight.
We considered only the \lya forest, excluding the region of the spectra bluewards of $\simeq$\SI{381.5}{nm}, and we masked both metal lines and saturated lines, to isolate the regions most sensitive to potential small-scale disturbances arising from stellar feedback \citep{Rauch01b}. 
We masked only metal lines that were surely identified from a redshift coincidence with absorption systems outside the \lya forest (i.e. lines from Table \ref{metal_list_A} and Table \ref{metal_list_B}). The regions to be masked were defined as the regions where the model of these lines falls below the continuum by more than 1 percent.
Saturated lines were masked by rejecting the regions where the model of \lya absorbers was lower than 0.05 in continuum units. The resulting usable portion of \lya forest amounts to a total wavelength range of \SI{61.48}{nm}.

Within this region we computed the normalized flux differences $\Delta f_{A-B}=f_A-f_B$ and $\Delta f_{B-A}=f_B-f_A$, and we look for absorption-like features in both these difference spectra. Absorption features were detected with the same prominence approach described in \autoref{Sect:LineList} and fitted with Voigt profiles using \AC{}.  We selected only features with $\log N_\mathrm{HI}>12$ and $b>\SI{5}{\km\per\second}$ that were sufficiently well fitted ($\chi_r^2<2$), to avoid including a handful of false detections (single-pixel residuals of data reduction mistakenly detected as lines). We visually inspected the detected features to exclude that they were contaminated by nearby metal lines whose tails might have escaped the masking procedure.

Only three features passed these criteria, 
all three along sightline A (i.e. detected on $\Delta f_{A-B}$).
They are shown in \autoref{Fig:diff_feats}. The collective rest equivalent width of these features is \SI{1e-2}{nm}, or 0.015 percent of the usable \lya forest, 
putting a stringent limit to the fraction of regions affected by local disturbances along the two sightlines.
An assessment of these features is uncertain,
as they barely extend beyond the 2-$\sigma$ contour obtained by propagating the local error on flux of $\Delta f_{A-B}$.

The identification of faint galaxies in close proximity to these features holds the potential to yield insights into feedback mechanisms. Future integral field observations, e.g. with the MUSE spectrograph \citep{MUSE2010}, may serve as a critical means to establish and elucidate this connection.
\section{Metal Lines}
\label{sec:metals}
\begin{figure}
    \centering
    \includegraphics[width=\linewidth]{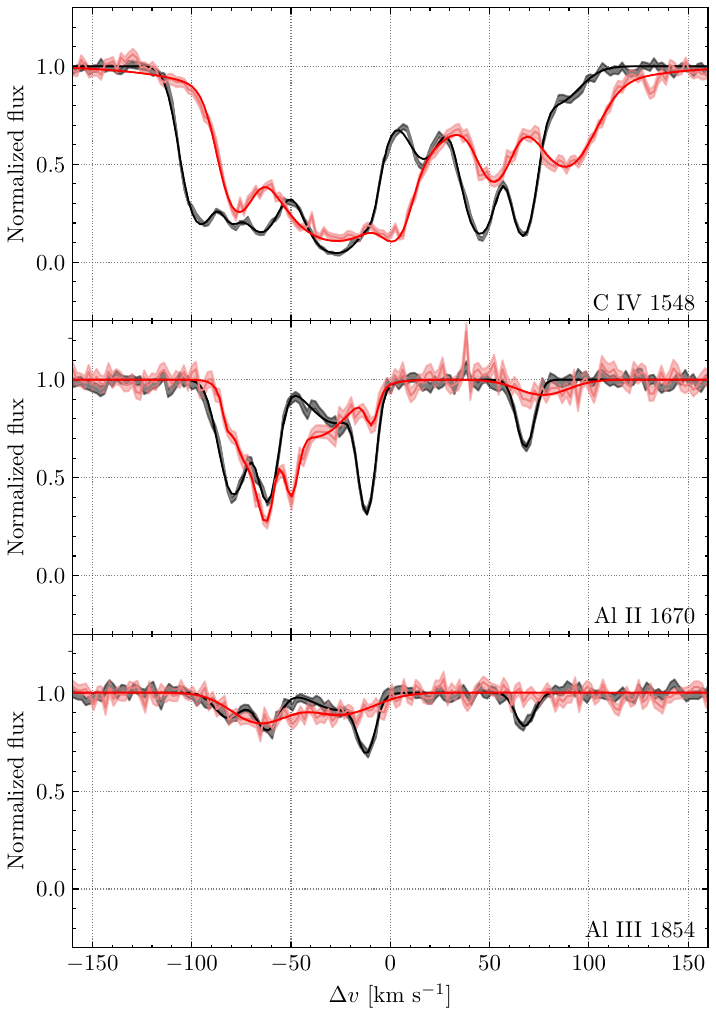}
    \caption{Detail of the C\textsc{iv}, Al\textsc{ii} and Al\textsc{iii} features of the metallic system at z=1.941 in velocity space. The solid lines define the absorption models of the two spectra. Fluxes are shown as a light-colored solid line, with a shaded area describing the noise. Image A is black; image B is red.}
    \label{fig:features_1.940}
\end{figure}

While the Lyman forests of the two images of UM673
appear remarkably similar, the metal lines show significant differences and a  global analysis like the one carried out in Section \ref{sect:global} would be neither effective nor unambiguous.
In order to study the differences and similarities of the metal absorptions along the two sightlines we need, first of all, an operational way to define an "individual metallic feature". Keeping in mind what is typically identified as a "feature" or a "complex of related components" by visual inspection, we have defined metallic absorption features as the spectral regions clearly bounded by two maxima in the flux transmission that can be unambiguously related to the same metallic transition. 
This definition is different from the one used for the \lya features in Section \ref{sect:lyafeat},
due to the inherent complexity of metal absorptions, which poses challenges in providing a comprehensive characterization.
Figure \ref{fig:features_1.940} shows some examples of these features.
The global traits of every feature can then be characterized by the following parameters:
\begin{enumerate}
    \item Ionic transition;
    \item Total column density, computed as the sum of the components' column densities;
    \item Redshift-space position, as a weighted mean of the components’ redshifts, with weights proportional to the measured column density;
    \item Equivalent width.
\end{enumerate}
We grouped together the features that have roughly the same redshift position and are likely to belong to the same physical structure, defining the so-called \textit{Metallic Systems}. Each system is characterized by the presence of one or more ionic transitions, whose velocity profiles resemble each other in shape and number of components. 
We report in Table \ref{tab:metfeat} the metallic systems visible in both images, their redshift as the mean of the values on the two sightlines and the transitions belonging to each system with their respective total column density. The H\textsc{i} column density is also reported for systems whose \lya falls within ESPRESSO's spectral range (i.e. $z>2.1$).
From the initial pool of all the features that were found, we selected the sample used for the rest of the analysis, based on the following criteria:
\begin{enumerate}
    \item A feature must be present in both sightlines and must have an equivalent width of at least 0.03 \AA\;  (observed). 
    \item A feature must not be contaminated by other transitions.
    \item A system must be at least $\SI{3000}{\km\per\s}$ away from the quasar emission redshift, in order to minimize the influence of the proximity effect and ensure the reliability of lens model. Effectively we chose systems with $z <2.68$.
\end{enumerate}
For ions that produce multiple transitions (e.g.  C\textsc{iv}, Fe\textsc{ii}), since all of them are modelled with the same parameters, only one feature has been considered. As a rule, we have picked up the strongest transition from the group or the feature that is not contaminated by other lines.
With these practical definitions, we defined a sample of 18  features belonging to 9 distinct metallic systems, reported in \autoref{tab:metfeat} with bold letters. Due to the second selection criterion, only three features fall within the Lyman forest. This fact reduces the size of our sample but also avoids systematic effects that would be difficult to address during the analysis. From the feature sample, we looked for significant relations amongst their parameters and their differences between the sightlines (e.g. $\Delta\log N = \log N_A-\log N_B$ vs. $\log N_{A,B}$, or  $\Delta EW=EW_A-EW_B $ vs. $z$). These relations are dominated by noise and no significant trend has been found. 

\begin{table}
\centering
  \begin{tabular}{lllllr} \toprule
    z & logN$_{H\textsc{i}}$ &Ion  & logN$_A$ & logN$_B$ & $\Delta v$ [km s$^{-1}$]  \\ \midrule
    0.564 & --- & \textbf{Mg\textsc{ii}}$^{\dagger}$ & 15.980 & 13.940 & 30.73 $\pm$ 0.66  \\
         && \textbf{Ca\textsc{ii}}& 11.808 & 12.134 & 32.18 $\pm$ 4.72  \\ 
         & &Fe\textsc{ii}$^{\dagger}$& 13.908 & 15.471  \\
    \midrule 
    1.357 & --- & \textbf{Mg\textsc{ii}}&  12.601 & 12.346 & 17.94 $\pm$ 0.84  \\
    \midrule
    1.626 & --- & Fe\textsc{ii} & 14.433 & 12.176 &  7.35 $\pm$ 1.45    \\
     &  & Mg\textsc{ii} & 15.169 & 12.048 & -7.23 $\pm$ 1.16 \\
          &  & C\textsc{iv}$^{\dagger}$& 13.250 & 13.405 \\
     \midrule 
      1.772 & --- & \textbf{C\textsc{iv}}$^{\dagger}$&  13.842 & 13.873 & 18.69 $\pm$ 1.55 \\
    \midrule 
      1.941  & --- & \textbf{C\textsc{ii}}$^\dagger$ & 14.756 & 14.990 & -2.44 $\pm$ 2.11 \\
      &&  \textbf{C\textsc{iv}}   & 14.539 & 14.470 &  -18.56 $\pm$ 0.67  \\
      && \textbf{Al\textsc{ii}}& 12.794 &  12.740 & -6.83 $\pm$ 1.74 \\     
      &&\textbf{Al\textsc{iii}}& 12.691  & 12.579 & -0.95 $\pm$ 3.68 \\
    && Si\textsc{iv}$^{\dagger}$ & 14.150 &  13.996  \\
    
    \midrule 
     1.944 &  --- &\textbf{C\textsc{iv}}& 14.239 & 13.834 & -1.59 $\pm$ 0.35 \\
     
    \midrule 
      2.060 &  --- &\textbf{C\textsc{iv}} & 13.236 & 12.915 & 1.77 $\pm$0.42 \\
      
    \midrule 
      2.066 & --- & \textbf{C\textsc{iv}} & 13.421 & 13.421 & -8.45 $\pm$ 2.61  \\
      
    \midrule 
     2.356 & 16.617 &  \textbf{Si\textsc{iv}} & 13.925 & 13.967 & 1.29 $\pm$ 1.62 \\
      && \textbf{Si\textsc{ii}}& 13.649 & 13.663 & 0.26 $\pm$ 0.28  \\
      && Si\textsc{iii}$^{\dagger}$& 14.201  &  15.194   \\
      && C\textsc{ii}$^{\dagger}$& 14.495 &  14.428  \\
      &&  \textbf{C\textsc{iv}} &  14.586 & 14.552 & -1.54 $\pm$ 0.42  \\
      && \textbf{Al\textsc{ii}} & 12.475 & 12.425 & 0.43 $\pm$ 0.23   \\
      && \textbf{Al\textsc{iii}}& 12.573 & 12.504 & -0.54 $\pm$ 0.34  \\
      
    \midrule 
     2.665 & 15.592 & \textbf{Si\textsc{iv}} & 11.892 &  12.131 & -1.02 $\pm$1.12 \\
     &&  \textbf{C\textsc{iv}} & 13.092 & 13.113 & -1.05 $\pm$ 0.69  \\

     \midrule
     2.707 & 15.014 &  C\textsc{iv} & 12.967& 12.891 & 1.18 $\pm$ 1.35 \\

     \midrule
     2.736 & 15.120 &  C\textsc{iv} & 13.230 & 13.149 & -0.44 $\pm$ 0.34  \\
    & &O\textsc{vi}& 13.791 &  16.639    \\
    
     \midrule
     2.739 & 13.285 &  C\textsc{iv} &12.631 & 12.556 & -0.66 $\pm$ 0.79  \\
    && O\textsc{vi}&14.148&  14.015 \\

    \bottomrule
  \end{tabular}
      \caption{List of all metallic systems that are detected on both sightlines. For each system we  indicate the average redshift position between the sightlines and the total H\textsc{i} column density (only for $z>2.1$ for which the \lya falls in ESPRESSO's range) and the total column density on each sightline. For some of the feature was possible to compute the velocity shift that maximizes $\xi_{CC}$. The $\dagger$ indicates transitions that fall within the Lyman forest, while features belonging to the final model sample (Sect.\ref{sec:metals}) are highlighted in bold letters.
}
\label{tab:metfeat}
  \end{table}

\subsection{Cross-Correlation Analysis}
\begin{figure}
    \centering
    \includegraphics[width=\linewidth]{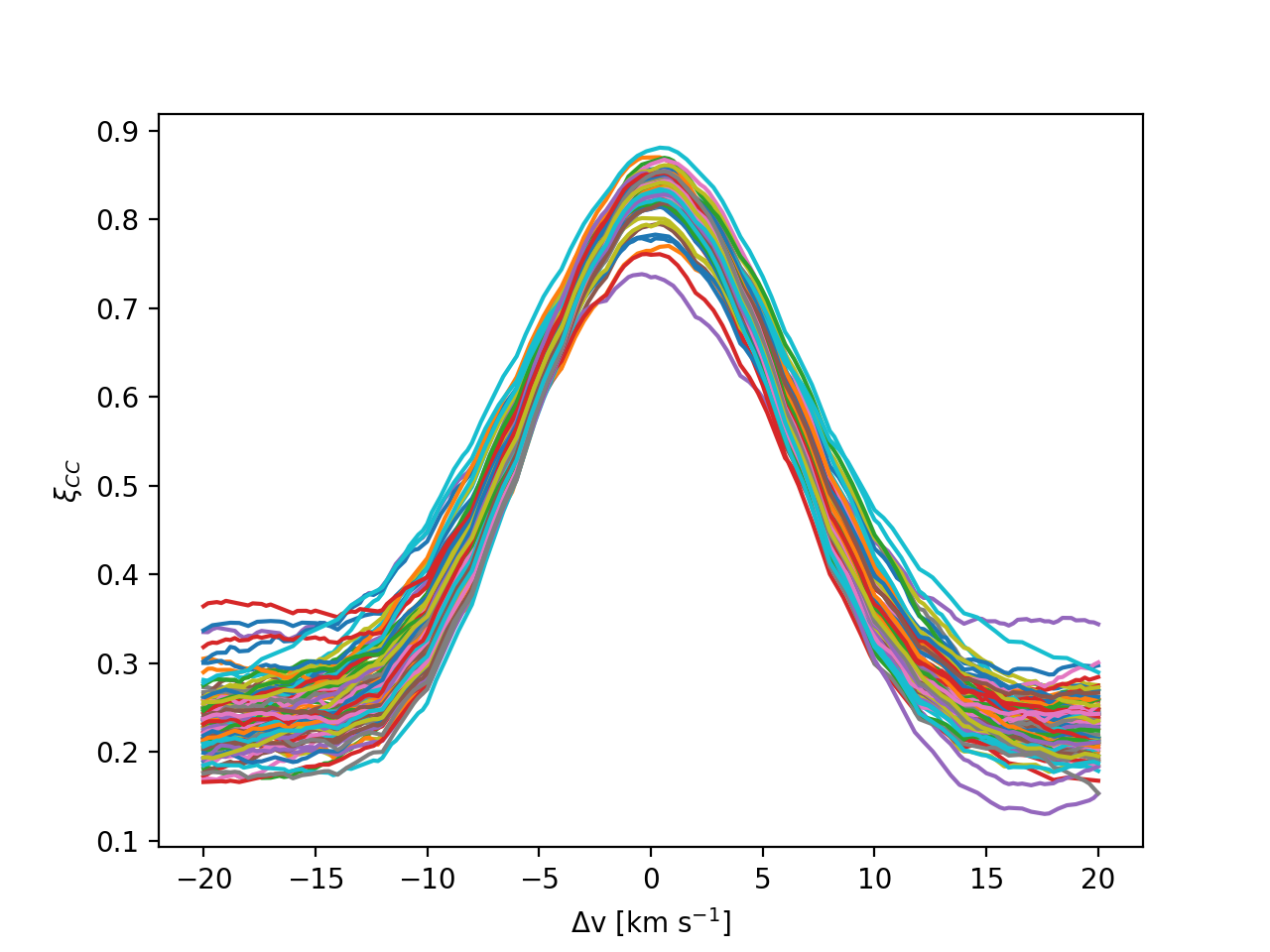}
    \caption{Cross-correlation function $\xi_{CC}$ profiles computed using the bootstrap method for the Al\textsc{ii} feature at $z = 2.356$. Every realization of $\xi_{CC}$ is shown with a different colour and has been fitted by a Gaussian profile. The distribution of the peaks position defines a velocity shift $\Delta v = 0.43\pm \SI{0.23}{\km\per\second}$.}
    \label{fig:ccf_alii}
\end{figure}

Similarly to the Lyman Forest, we computed the CCF for every metallic feature in the selected sample, highlighted with bold characters in Table \ref{tab:metfeat} as a function of velocity lag between the sightlines. 
As the metal features differ significantly in the two spectra
and the distribution of the residuals is expected to be non-gaussian, we have adopted a bootstrap method \citep{BOOTSTRAP}, which is useful when the theoretical distribution of a statistic of interest is complex or unknown, in order to estimate the position error of the peak of the CCF,  as shown in Figure \ref{fig:ccf_alii}. 
The method is similar to the one described by \cite{Peterson1998}. Using \AC{}, we created an ensemble of 100 realizations for the two sightlines by randomly sampling the pairs of flux values from the observed spectra. The sampling was done with replacement, but counting just once the pairs that were selected multiple times; due to this approach, the typical number of spectral bins in the realizations was a factor approximately $1-1/e$ smaller than the number of spectral bins in the observed spectra. For each pair of realizations in our ensemble, we computed the CCF and fitted a Gaussian profile to its core to determine the position of the maximum.
The resulting $\Delta v$ that maximize the CCF are reported in Table \ref{tab:metfeat}.

For each system, we can estimate an average redshift position, which directly relates to a separation between the sightlines given by the lens models (see Section \ref{Sect:LensModel}).
In order to take into account the uncertainties related to the lens model, we assume for every redshift position a sightline separation given by
$
    l(z) =( l_{3}(z)+l_{1}(z))/2,
$
where $l_1$ and $l_3$ are the separations given by Model 1 and Model 3 by \cite{Koptelova14}, respectively. Then, the uncertainty in the separation  has been computed as
$
    \sigma_l(z) = (l_{3}(z)-l_{1}(z))/2.
$
The redshift uncertainty of the systems, being negligible, has not been considered.

Figure~ \ref{fig:halo_model} shows the absolute value of the velocity lag that maximises the CCF for each feature as a function of sightline separation.
In the figure, the velocity difference is below $ \SI{2}{\km\per\second}$ for transversal separations $\lesssim 800$ pc, and increases up to $\sim \SI{35}{\km\per\second}$ at higher separations. The observational pattern indicates that the morphology of high-redshift metal absorbers lacks discernible features at scales below a few hundred parsecs, whereas notable variations in the absorption characteristics become evident when examining separations on the scale of kiloparsecs.
Besides, it is plausible that this relationship
extends above the 10 kpc scale probed by our observation and thus should be investigated with different lenses or pairs of QSOs in future studies.
\subsection{Characterization of the metal absorbers} \label{sect:toymodel}
To investigate the implications of the metal analysis results, we devised a toy model designed to simulate the absorption of two sightlines by a gaseous structure. 
It should be noted that the primary objective of our simplified approach was to gain insights into the typical environments responsible for the observed trend and provide an order-of-magnitude estimation of their mass and dimensions. 
Future analyses, in particular when more data will be available on other lensed systems at different redshifts and separations, will require cosmological simulations \cite[e.g.][]{Shin2021} to better capture the complexity of the astrophysical processes governing galactic feedback in the CGM and IGM, along with fluctuations in the ionizing UVB due to HeII reionization.

From the absorption line list obtained in  Section \ref{Sect:LineList}, we computed the total H\textsc{i} column densities associated with the selected metallic systems and found values within $13.2 < \log\left( N_{\rm {HI}}/{\rm cm^{-2}} \right)< 17$. The system at $z=1.626$, as described by \citet{Cooke2010}, belongs to a Damped \lya System and has not been considered in the subsequent analysis, due to its low redshift that places most of its metal lines in the Lyman forest.

The typical H\textsc{i} column densities of the systems listed in Table~\ref{tab:metfeat} suggest that they do not belong to the general IGM, but to galactic halo structures or to the circumgalactic medium.
This is corroborated by the metallicity of the system at $z=2.356$, the only one with a number of transitions that allows a comprehensive analysis with the CLOUDY package \citep{CLOUDY2017}. Its metallicity is found to range between solar and one-tenth solar, primarily depending on the assumptions about the UV ionizing background \citep{HaardtMadau1996, HaardtMadau12}.

\subsubsection{A Toy Halo Model}
We developed a simple toy-model of a DM halo hosting the absorbing structure, defined by a virial mass $M$ that we assumed constant in redshift as a zeroth-order approximation. 
In a $\Lambda$CDM cosmology a virialized halo has an average density given by \citep{mo2010galaxy}
\begin{equation}
         \overline{\rho_h}(z) = \Delta_v(z)~\Omega_m(z)\frac{3H^2(z)}{8\pi G},
\end{equation}
where the virial overdensity $\Delta_v$ can be approximated by \citep{BryanNorman98}
 \begin{equation}\label{eq:Delta}
     \Delta_v(z) = \frac{18\pi^2+82(\Omega_m(z)-1)-39(\Omega_m(z)-1)^2}{\Omega_m(z)}.
 \end{equation}
 The halo's virial radius has been computed as the radius of a sphere of mass $M$ and average density $\overline{\rho_h}$. We assumed a NFW circular velocity profile \citep{Navarro96}
 \begin{equation}\label{Eq:velprof}
    V_h(r,z) = V_{vir}(z)\cdot\sqrt{\frac{f(c\cdot r/R_{vir}(z))}{r/R_{vir}(z)\cdot f(c)}} ,
\end{equation}
where 
\begin{equation}
   f(x) = \ln (x+1)-x/(x+1),
\end{equation}
while $V_{vir}$ is defined as the virial velocity
and $c\sim 5$ is the typical concentration parameter for a dark-matter halo at $z\sim2$ \citep{Zhao2009}.
Since we are interested in the velocity of the baryonic component of the halo, we add a factor 1.5 in Eq. \ref{Eq:velprof} to account for the rotation of the baryons that, as numerical simulations show \citep{mo2010galaxy}, spin about 1.5 times faster than the hosting DM. 
We used relation 9 of \cite{ValeOstriker06} to link halo mass to the luminosity of the hosted absorbing structure and computed the radius of the absorber through the Holmberg scaling relation \citep{Holmberg75}
\begin{equation}\label{holmberg}
    R_g = R_\star\left(\frac{L}{L_\star}\right)^\beta,
\end{equation}
where $\beta=0.4$ \citep{Chen2001,Nielsen2013} 
and $R_\star$ is the effective absorbing radius of a typical $L_\star$ galaxy, that for absorption features of the typical EW limit of our observation is assumed to be $R_\star=250$ kpc, linearly extrapolating the results of \cite{Hasan2020}. As a first order approximation we neglected the redshift evolution of $R_\star$.
Finally,
we pierced the halo model with 1000 sightline-pairs 
separated by $l_i$, with $0.1 < \log l_i < 1.1$ ($l_i$ in units of kpc). The position of each sightline-pair is randomly
chosen from a uniform probability circle of radius $R_g$ around
the projected sky position of the halo center.
The separation $l_i$ defines the halo's redshift through the lens model. 
Every pair pierces a different halo with a randomly oriented spin.
For the sake of simplicity, we assumed that the velocity of an absorption feature is related to the point along the sightline closest to the halo center. Therefore, we computed the projected velocity of the baryonic matter along the sightlines at the closest point to the halo center and retrieve the distribution of the velocity differences between the pairs as a function of the sightline separation. 

\subsubsection{Best-fit Model}
\begin{figure}
    \centering
    \includegraphics[width=\linewidth]{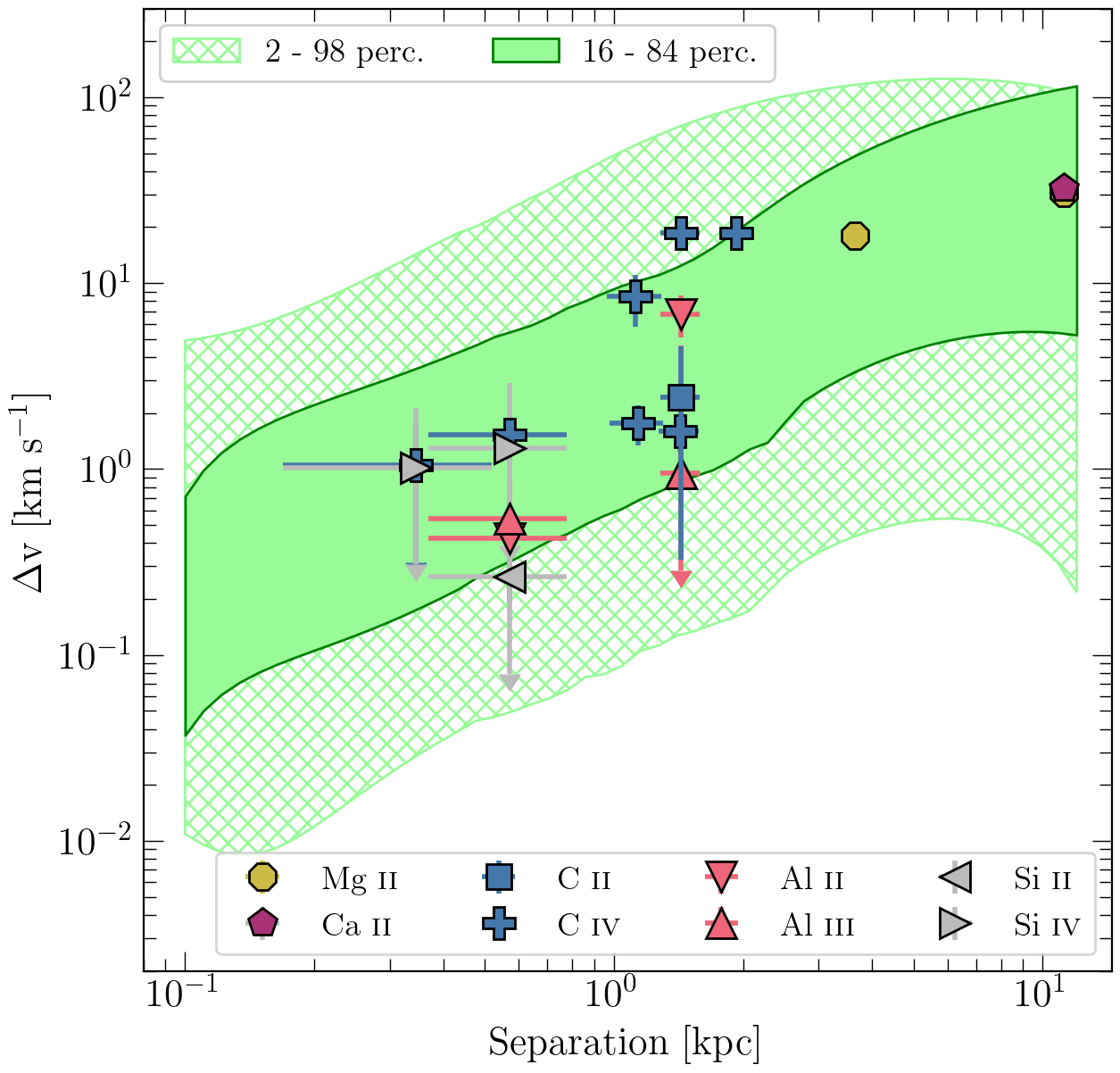}
    \caption{
    Velocity shifts of metallic spectral features versus the separation between sightlines. The absolute values of the velocity shifts, determined through CCF analysis, are displayed for all metal features within the sample, with distinct ions represented in different colors. Areas shaded in green denote the smoothed 16-84 and 2-98 percentile intervals of the model's velocity distribution that better fits the C\textsc{iv} absorbers. This model is derived using the best-fit mass of $M=1.9 \times 10^{10} M_\odot$. To enhance visual clarity, instances of low-separation systems exhibiting velocity shifts $\Delta v$ that are consistent with zero are represented as upper limits.}
    \label{fig:halo_model}
\end{figure}

We compared the halo model results to the data in order to estimate the typical halo mass that gives rise to the observed absorption trend. 
For consistency, we only used the seven C\textsc{iv} absorbers with $| \Delta v |  > 3000\ \kms$ with respect to the emission redshift (blue crosses in Figure \ref{fig:halo_model}) to constraint the halo model. 
Given a mass value, we estimated the probability of such model to recreate the observed data points by computing the difference between the percentiles of the model-generated velocity distribution that fall between $\Delta v_i \pm \sigma_i$ for every observed data point. The product of the percentile differences on all the data points gives an estimate on the probability of such mass to recreate the observed trend. 

The probability estimate has been computed on 50 realizations of the velocity distribution for 50 log-scaled mass values within $5\times10^9\leq M/M_\odot\leq 10^{12}$. The average of all 50 runs has been smoothed over a 2-bins wide Gaussian window and peaks at $M=1.9\times 10^{10} M_\odot$. Figure \ref{fig:halo_model} shows the 16-86 and 2-98 percentile intervals of the velocity difference distribution computed by the model assuming the best-fitting mass value. The C\textsc{iv} systems used to constrain the model as well as the other features from the metal sample are shown. 

\subsubsection{Cosmic Incidence}
The consistency of the previous mass estimate can be roughly checked by computing the halo's cosmic incidence (i.e. the typical number of halos pierced by a sightline per unit of comoving length) and comparing it to the number of C\textsc{iv} features found on the spectra that fall redwards of the \lya emission. We found $N=6$ systems over a redshift range $\Delta z=0.724$ ($1.941<z<2.665$). We can estimate the comoving path density of C\textsc{iv} absorbers as
\begin{equation}
    \frac{dN}{dX}=\frac{dN}{dz}\frac{dz}{dX}\approx\frac{N}{\Delta z}\frac{dz}{dX},
\end{equation}
where
\begin{equation}
\frac{dX}{dz}=\frac{(1+z)^2}{\sqrt{\Omega_{m}(1+z)^3+\Omega_{\Lambda}}}.
\end{equation}
At $z\sim2$ this estimate yields $dN/dX\sim 2.67$. 

On the other hand, the cosmic incidence of our typical halo structure can be computed as
\begin{equation}
    \frac{dN}{dX} = \frac{c}{H_0}n\sigma,
\end{equation}
where $\sigma=\pi R_g^2$ is the spherical cross-sectional area of the absorbing gas (assuming a unity covering fraction $f_c=1$). The cosmic number density of gas structures $n$ can be obtained by integrating the galaxy luminosity function derived from \cite{Parsa2016}
\begin{equation}
    \Phi(L)dL = \Phi_\star\left(\frac{L}{L_\star}\right)^\alpha\exp\left(\frac{L}{L_\star}\right)\frac{dL}{L_\star},
\end{equation}
with parameters, evaluated at $z = 2$, $\Phi_\star = 0.002$, $\alpha = -1.4$ and $L_\star = 2.2\times10^{10} L_\odot$, as fitted from observed UV luminosity functions over the range $0\le z \le 8$. We integrated the product $n\sigma$ over luminosity from a minimum luminosity $L_{min}$ and retrieved the halo cosmic incidence at
$z =2$.
\begin{equation}
    \frac{dN}{dX} = \frac{\pi c \Phi_\star}{H_0}\Gamma[x,l]R_\star^2,
\end{equation}
where $\Gamma[x,l]$ is the upper incomplete gamma function
\begin{equation}
    \Gamma[x,l] = \int_l^\infty t^{x-1}e^{-t}dt,
\end{equation}
while $l = L_{min}/L_\star$ and $x=2\beta+\alpha+1$.
Assuming that the lower luminosity bound of the integration $L_{min}/L_{\star}$ is related to the estimated halo mass $M = 1.9\times10^{10} M_\odot$ through relation 9 of \cite{ValeOstriker06}, we compute the halo's cosmic incidence $dN/dX = 3.73$, compatible with the C\textsc{iv} estimate. 

\section{Discussion and Conclusions}
In this work we have shown the potential provided by the spectroscopic analysis of lensed quasars in the context of the study of the dynamical properties of the IGM at different scales. We provided proof of the capabilities of the ESPRESSO spectrograph in this field, that can be exploited by the flexible analysis pipeline of the {\textbf\AC{}} environment. Here we outline the results of our analysis on the small-scale structure of the IGM.

\subsection{Lyman forest}
The analysis of the \lya features highlighted the absence of any global systematic drift effects between the two sightlines and showed that the width of the CCF profile is due to the width of the lines and the typical clustering scale of the \lya forest in velocity space. With a feature-by-feature approach, we also showed that the \lya clouds motions are coherent on sub-kpc scales with a confidence level of $\sim \SI{1}{\km\per\s}$.

We investigated if with the present observational setup 
the effects of the universal expansion on
the \lya absorbers, modeled as slabs following the expansion,
could be detected.
The expected median $\Delta v \sim\SI{120}{\m\per\second}$, for a typical separation of \SI{0.5}{kpc} at a $\langle z\rangle \sim 2.45$, turns out to be below the sensitivity of the present observations,
but a detection is not beyond reach, if a higher SNR or/and a wider separation of sightlines is achieved.

In the context of the Sandage test, the observed global shift of the Lyman forest of $\Delta v = 12 \pm \SI{48}{\m\per\s}$ can be compared with the velocity accuracy scaling relation predicted by \cite{Liske08}:
\begin{equation}
    \sigma_v = 2\left(\frac{SNR}{2370}\right)^{-1}\left(\frac{N_{QSO}}{30}\right)^{-1/2}\left(\frac{1+z_{QSO}}{5}\right)^{-1.7}\;\text{cm s}^{-1},
\end{equation}
where $SNR$ is the average signal-to-noise ratio per $0.00125$ nm pixel of the \lya forest and $N_{QSO}$ is the number of spectra of the same source. In our case, images A and B must be considered as separated sources and, with an average SNR of $\sim 26$ and $\sim 16$, respectively, per $\sim 0.0028$ nm pixel in the Lyman forest ($\sim 90$\%
of which has been used for the CCF), we obtain an expected total accuracy in the velocity position of the features
\begin{equation}
    \sigma_v = \sqrt{\sigma_{v,A}^2+\sigma^2_{v,B}}\simeq \SI{50}{\m\per\s}
\end{equation}
which is the same order of magnitude uncertainty recovered by the CCF process developed in Section \ref{sect:global}, corroborating the proposed scaling relation on real data, even at a relatively low SNR.

Lastly, we can check whether the infinitesimal changes in the \lya absorbers over a given separation will produce systematic effect in the measure of the redshift drift. From \cite{Dodorico2006} we assume that the \lya clouds have peculiar velocities likely smaller than $\lesssim \SI{100}{\km\per\second}$. Over a 20 years baseline for an hypothetical redshift drift experiment, these clouds would cover a distance of about 0.002 pc in proper units. In Section \ref{sect:lyafeat} we observed a null velocity shift of the \lya features on 1 kpc scales, with an upper bound of $\SI{1}{\km\per\second}$. Linearly extrapolating this upper bound to milliparsec scales, we retrieve a velocity lag of $\sim \SI{0.2}{\cm\per\second}$, which is about two orders of magnitude smaller than the expected redshift drift effect of $\SI{12}{\cm\per\second}$ over the same 20 years baseline. Therefore, the peculiar velocities of the \lya forest appear to be too weak to induce significant effects on experiments aimed at measuring $\dot{z}$.

\subsection{Metals}
In contrast to the \lya clouds, significant differences are observed in the metallic absorption pattern between the two images. We conducted an analysis of these dissimilarities utilizing the cross-correlation function of the fluxes as a function of velocity lag and evidenced a relationship between the observed velocity shift along the sightlines and the spatial separations of the metallic absorbers.

We have observed that the velocity lag of these absorbers remains close to zero for scales below $800$ pc. Although one might be tempted to infer coherence among the metal absorbers below this scale, our sample lacks strong metallic absorption instances at $z>2.5$ (the typical redshifts sampling shorter separation scales). Consequently, the higher-redshift features with shorter separations may appear indistinguishable within the noise, primarily due to their relatively small column densities.
Through the use of a toy model, we have shown how the relation between velocity difference ($\Delta v$) and separation can be induced by rotating virialized dark matter halos with a virial mass of approximately $M_{\text{vir}} \sim 1.9\times 10^{10}M_\odot$. It is important to note that our analysis provides only an order of magnitude estimate. Nonetheless, even with this rudimentary model, we successfully replicated the observed trend and found consistency with the cosmic incidence of the absorbers.

In relation to the Sandage test of the cosmological redshift drift, considering the halo model discussed above, it can be inferred that the metallic absorbers exhibit peculiar velocities of the order of $\sim \SI{200}{\km\per\second}$. Consequently, over a duration of $20$ years, these absorbers are anticipated to traverse a proper distance of approximately $0.004$ pc. By linearly scaling the observed velocity shear of $\Delta v \sim \SI{10}{\km\per\second}$ at $1$ kpc to the $0.004$ pc scale, we can estimate the expected velocity change of a metallic absorber over a 20-year interval, which amounts to $\sim \SI{4}{\cm\per\second}$.
The velocity difference resulting from the motion of metallic absorbers over a 20-year baseline is thus of the same order of magnitude as the velocity shift induced by the accelerated expansion of the Universe at the typical redshifts considered for the test. These velocity differences introduce a non-negligible noise effect that needs to be taken into account when measuring $\dot{z}$ using metallic lines. Moreover, this noise effect may pose a  substantial limitation for specific approaches such as the \textit{\lya cell} technique proposed by \citet{Cooke2020}.  

The sample of metallic features used in this pilot study has a limited size, but the results are promising and intriguing:  future developments will focus on a broader range of lenses, encompassing varying separations and redshifts, to validate the $\Delta v-\text{Separation}$ relationship, ascertain its evolution with redshift, and explore potential dependencies on the ionization state of the absorbers.
Increasing the SNR of of both the existing and forthcoming data sets will yield a larger number of absorbers while reducing measurement uncertainties. This, in turn, will provide more stringent constraints and illuminate the nature of the absorbing structures to a greater extent.

\section*{Acknowledgements}
The authors acknowledge the ESPRESSO project team for its effort and dedication in building the ESPRESSO instrument.
The INAF authors acknowledge financial support of the Italian Ministry of Education, University, and Research with PRIN 201278X4FL and the "Progetti Premiali" funding scheme. 
JIGH acknowledges financial support from the Spanish Ministry of Science and Innovation (MICINN) project PID2020-117493GB-I00.
SC and DM are partly supported by the INFN PD51 INDARK grant.
This work was financed by Portuguese funds through FCT - Funda\c c\~ao para a Ci\^encia e a Tecnologia through national funds and in the framework of the project 2022.04048.PTDC,
and by FEDER through COMPETE2020 - Programa Operacional Competitividade e Internacionalização by these grants: UIDB/04434/2020; UIDP/04434/2020.
SGS acknowledges the support from FCT through Investigador FCT contract nr. CEECIND/00826/2018 and  POPH/FSE (EC).
CJM also acknowledges FCT and POCH/FSE (EC) support through Investigador FCT Contract 2021.01214.CEECIND/CP1658/CT0001. 
FPE and CLO would like to acknowledge the Swiss National Science Foundation (SNSF) for supporting research with ESPRESSO through the SNSF grants nr. 140649, 152721, 166227 and 184618. The ESPRESSO Instrument Project was partially funded through SNSF’s FLARE Programme for large infrastructures. 
TMS acknowledgment the support from the SNF synergia grant CRSII5-193689 (BLUVES).
MTM acknowledges the support of the Australian Research Council through Future Fellowship grant FT180100194

\section*{Data Availability}
The data underlying this article will be shared on reasonable request to the corresponding author.



\bibliographystyle{mnras}
\bibliography{00ESP_LensSC} 



\onecolumn

\begin{longtable}{ll ll ll ll l}
	\caption{Lyman absorption-line parameter list - complete version in electronic form}\\
    \label{lya_list}\\
	\toprule
	\multicolumn{1}{c}{$\lambda_{obs}$} & \multicolumn{1}{c}{Ion} & \multicolumn{1}{c}{$\lambda_{rest}$} & \multicolumn{1}{c}{z} & \multicolumn{1}{c}{$\sigma_z$} & \multicolumn{1}{c}{logN} & \multicolumn{1}{c}{$\sigma_{\text{logN}}$} & \multicolumn{1}{c}{b} & \multicolumn{1}{c}{$\sigma_{\text{b}}$} \\
	\multicolumn{1}{c}{nm} & & \multicolumn{1}{c}{nm} & & & & & \multicolumn{1}{c}{km s$^{-1}$} & \multicolumn{1}{c}{km s$^{-1}$} \\
	\midrule
	\endfirsthead
	\caption{Lyman absorption-line parameter list (continued)}\\
	\toprule
	\multicolumn{1}{c}{$\lambda_{obs}$} & \multicolumn{1}{c}{Ion} & \multicolumn{1}{c}{$\lambda_{rest}$} & \multicolumn{1}{c}{z} & \multicolumn{1}{c}{$\sigma_z$} & \multicolumn{1}{c}{logN} & \multicolumn{1}{c}{$\sigma_{\text{logN}}$} & \multicolumn{1}{c}{b} & \multicolumn{1}{c}{$\sigma_{\text{b}}$} \\
	\multicolumn{1}{c}{nm} & & \multicolumn{1}{c}{nm} & & & & & \multicolumn{1}{c}{km $s^{-1}$} & \multicolumn{1}{c}{km $s^{-1}$} \\
	\midrule
	\endhead
	\midrule 
	\multicolumn{9}{r}{\footnotesize\itshape Continue on the next page}
	\endfoot
	\bottomrule
	\endlastfoot
380.0209 & Ly  $\beta$& 102.5722 & 2.70491 & 0.00001 & 13.594 & 0.005 & 28.31 & 0.38  \\
380.0862 & Ly $\alpha$ & 121.5670 & 2.12656 & 0.00016 & 13.115 & 1.051 & 12.19 & 10.62 \\
380.1044 & Ly $\beta$ & 102.5722 & 2.70572 & 0.00001 & 13.550 & 0.016 & 23.55 & 0.67  \\
380.1643 & Ly $\alpha$ & 121.5670 & 2.12720 & 0.00015 & 15.227 & 2.662 & 26.66 & 27.15 \\
380.2085 & Ly $\beta$ & 102.5722 & 2.70674 & 0.00028 & 14.988 & 0.601 & 30.70 & 7.23  \\
\end{longtable}

\begin{longtable}{lllrrrrll}
	\caption{Rest-frame equivalent width, velocity shift, and total column density for the most reliable \lya features}  \\
\toprule
\multicolumn{1}{c}{$\lambda_{obs}$} & \multicolumn{1}{c}{EW$_A$} & \multicolumn{1}{c}{EW$_B$} & \multicolumn{1}{c}{$\Delta v_A$} & \multicolumn{1}{c}{$\Delta v_B$} & \multicolumn{1}{c}{$\Delta v_A - \Delta v_B$}  & \multicolumn{1}{c}{$\log N_\mathrm{HI}^A$} & \multicolumn{1}{c}{$\log N_\mathrm{HI}^B$} \\
\multicolumn{1}{c}{nm} & \multicolumn{1}{c}{nm}&\multicolumn{1}{c}{nm} &\multicolumn{1}{r}{km s$^{-1}$} & \multicolumn{1}{r}{km s$^{-1}$}& \multicolumn{1}{c}{km s$^{-1}$} & \multicolumn{1}{c}{$N_\mathrm{HI}$ in cm$^{-2}$}  & \multicolumn{1}{c}{$N_\mathrm{HI}$ in cm$^{-2}$}\\
\midrule
\endfirsthead
\caption{Equivalent width, velocity shift, and total column density for the most reliable \lya features (continued)}\\
\toprule
\multicolumn{1}{c}{$\lambda_{obs}$} & \multicolumn{1}{c}{EW$_A$} & \multicolumn{1}{c}{EW$_B$} & \multicolumn{1}{c}{$\Delta v_A$} & \multicolumn{1}{c}{$\Delta v_B$} & \multicolumn{1}{c}{$\Delta v_A - \Delta v_B$} & \multicolumn{1}{c}{$\log N_\mathrm{HI}^A$} & \multicolumn{1}{c}{$\log N_\mathrm{HI}^B$}\\
\multicolumn{1}{c}{nm} & \multicolumn{1}{c}{nm}&\multicolumn{1}{c}{nm} &\multicolumn{1}{r}{km s$^{-1}$} & \multicolumn{1}{r}{km s$^{-1}$}& \multicolumn{1}{c}{km s$^{-1}$} & \multicolumn{1}{c}{$N_\mathrm{HI}$ in cm$^{-2}$} & \multicolumn{1}{c}{$N_\mathrm{HI}$ in cm$^{-2}$}\\
\midrule

\endhead
\bottomrule
\multicolumn{9}{r}{\footnotesize\itshape Continue on the next page}
\endfoot
\endlastfoot

381.3841 & 0.0187 $\pm$ 0.0012 & 0.0145 $\pm$ 0.0026 & -2.6439 & 0.5768  & -3.2208 & 14.2561 $\pm$ 0.0045 & 14.2520 $\pm$ 0.0102 \\
384.3842 & 0.0131 $\pm$ 0.0006 & 0.0144 $\pm$ 0.0013 & 1.2231  & 3.0422  & -1.8191 & 13.5596 $\pm$ 0.0142 & 13.6149 $\pm$ 0.0429 \\
386.7197 & 0.0214 $\pm$ 0.0006 & 0.0188 $\pm$ 0.0015 & 0.3764  & -3.5330 & 3.9094  & 13.8300 $\pm$ 0.0108 & 13.7970 $\pm$ 0.0391 \\
386.8926 & 0.0296 $\pm$ 0.0006 & 0.0285 $\pm$ 0.0014 & -0.0055 & -0.3560 & 0.3506  & 14.1487 $\pm$ 0.0147 & 14.1065 $\pm$ 0.0625 \\
387.2102 & 0.0179 $\pm$ 0.0005 & 0.0174 $\pm$ 0.0012 & -0.3713 & -2.1986 & 1.8273  & 13.7324 $\pm$ 0.0123 & 13.6982 $\pm$ 0.0364 \\
391.0653 & 0.0489 $\pm$ 0.0008 & 0.0513 $\pm$ 0.0017 & -0.1023 & 0.2954  & -0.3977 & 16.9554 $\pm$ 0.0598 & 16.9572 $\pm$ 0.1846 \\
392.0383 & 0.0152 $\pm$ 0.0004 & 0.0120 $\pm$ 0.0010 & -0.1435 & -0.7871 & 0.6436  & 13.7056 $\pm$ 0.0135 & 13.6929 $\pm$ 0.0469 \\
393.9732 & 0.0332 $\pm$ 0.0005 & 0.0313 $\pm$ 0.0011 & 0.1528  & 2.3562  & -2.2034 & 14.3705 $\pm$ 0.0185 & 14.3551 $\pm$ 0.0641 \\
395.4676 & 0.0321 $\pm$ 0.0003 & 0.0310 $\pm$ 0.0008 & -0.0332 & -0.4598 & 0.4266  & 14.3997 $\pm$ 0.0131 & 14.3321 $\pm$ 0.0348 \\
397.4632 & 0.0157 $\pm$ 0.0003 & 0.0144 $\pm$ 0.0007 & 0.7912  & -1.6025 & 2.3936  & 13.6561 $\pm$ 0.0107 & 13.6111 $\pm$ 0.0251 \\
398.1970  & 0.0193 $\pm$ 0.0004 & 0.0160 $\pm$ 0.0007 & 0.3577  & 1.6928  & -1.3352 & 13.6531 $\pm$ 0.0079 & 13.6118 $\pm$ 0.0284 \\
399.1399 & 0.0335 $\pm$ 0.0003 & 0.0350 $\pm$ 0.0006 & -0.3288 & -2.3967 & 2.0679  & 14.1741 $\pm$ 0.0101 & 14.1648 $\pm$ 0.0250 \\
402.7939 & 0.0222 $\pm$ 0.0003 & 0.0231 $\pm$ 0.0005 & -0.3539 & -1.8350 & 1.4811  & 13.8225 $\pm$ 0.0054 & 13.8180 $\pm$ 0.1049 \\
403.9995 & 0.0369 $\pm$ 0.0003 & 0.0354 $\pm$ 0.0005 & -0.4680 & -1.9330 & 1.4650  & 14.3742 $\pm$ 0.0297 & 14.3161 $\pm$ 0.0489 \\
405.7809 & 0.0167 $\pm$ 0.0002 & 0.0157 $\pm$ 0.0003 & -0.1363 & -0.5177 & 0.3814  & 13.7777 $\pm$ 0.0072 & 13.7468 $\pm$ 0.0186 \\
408.1144 & 0.1801 $\pm$ 0.0004 & 0.1755 $\pm$ 0.0010 & 4.6080  & 8.0095  & -3.4015 & 16.6175 $\pm$ 0.0221 & 16.6175 $\pm$ 0.0011 \\
409.2676 & 0.0416 $\pm$ 0.0003 & 0.0401 $\pm$ 0.0005 & -0.3019 & 1.9445  & -2.2464 & 14.4354 $\pm$ 0.0425 & 14.4354 $\pm$ 0.0011 \\
412.1473 & 0.0243 $\pm$ 0.0002 & 0.0233 $\pm$ 0.0004 & -0.0814 & -0.3395 & 0.2581  & 13.8506 $\pm$ 0.0183 & 13.8394 $\pm$ 0.0240 \\
412.3095 & 0.0201 $\pm$ 0.0002 & 0.0202 $\pm$ 0.0004 & 0.2034  & 1.2999  & -1.0966 & 13.7727 $\pm$ 0.0175 & 13.7701 $\pm$ 0.0215 \\
415.7067 & 0.0374 $\pm$ 0.0003 & 0.0372 $\pm$ 0.0006 & -0.0939 & -0.8630 & 0.7691  & 14.5136 $\pm$ 0.0088 & 14.5164 $\pm$ 0.0250 \\
419.3809 & 0.0189 $\pm$ 0.0002 & 0.0183 $\pm$ 0.0003 & 0.3584  & 1.5384  & -1.1800 & 13.7664 $\pm$ 0.0044 & 13.7639 $\pm$ 0.0132 \\
419.6957 & 0.0143 $\pm$ 0.0002 & 0.0138 $\pm$ 0.0003 & 1.0043  & 0.6311  & 0.3732  & 13.5951 $\pm$ 0.0045 & 13.6061 $\pm$ 0.0124 \\
424.4629 & 0.0540 $\pm$ 0.0002 & 0.0531 $\pm$ 0.0004 & -0.0595 & -1.4489 & 1.3894  & 14.4509 $\pm$ 0.0063 & 14.4522 $\pm$ 0.0192 \\
425.4665 & 0.0518 $\pm$ 0.0002 & 0.0512 $\pm$ 0.0004 & 0.5152  & 0.8238  & -0.3085 & 14.9767 $\pm$ 0.0079 & 14.9791 $\pm$ 0.0229 \\
427.8303 & 0.0174 $\pm$ 0.0002 & 0.0169 $\pm$ 0.0003 & 0.2240  & -0.4698 & 0.6938  & 13.7173 $\pm$ 0.0064 & 13.6916 $\pm$ 0.0137 \\
429.6981 & 0.0163 $\pm$ 0.0002 & 0.0151 $\pm$ 0.0003 & -0.8160 & -1.8199 & 1.0039  & 13.1523 $\pm$ 0.0300 & 13.1745 $\pm$ 0.0564 \\
431.2187 & 0.0100 $\pm$ 0.0001 & 0.0093 $\pm$ 0.0003 & 0.3789  & 0.2675  & 0.1113  & 13.3923 $\pm$ 0.0078 & 13.3448 $\pm$ 0.0173 \\
433.0306 & 0.0090 $\pm$ 0.0002 & 0.0093 $\pm$ 0.0003 & -0.1111 & -0.7316 & 0.6205  & 13.3046 $\pm$ 0.0062 & 13.3506 $\pm$ 0.0144 \\
433.6420  & 0.0107 $\pm$ 0.0002 & 0.0105 $\pm$ 0.0003 & -0.2101 & 0.5687  & -0.7789 & 13.4160 $\pm$ 0.0054 & 13.4121 $\pm$ 0.0157 \\
433.7606 & 0.0101 $\pm$ 0.0001 & 0.0096 $\pm$ 0.0003 & -0.3579 & 1.0194  & -1.3773 & 13.4169 $\pm$ 0.0051 & 13.3471 $\pm$ 0.0163 \\
434.1963 & 0.0166 $\pm$ 0.0001 & 0.0161 $\pm$ 0.0003 & -1.3570 & -0.7051 & -0.6519 & 13.7457 $\pm$ 0.0044 & 13.7264 $\pm$ 0.0116 \\
436.1734 & 0.0540 $\pm$ 0.0001 & 0.0524 $\pm$ 0.0003 & -0.0387 & 0.3581  & -0.3968 & 15.1151 $\pm$ 0.0067 & 15.0983 $\pm$ 0.0154 \\
436.6626 & 0.0427 $\pm$ 0.0002 & 0.0420 $\pm$ 0.0003 & -0.9607 & -0.7653 & -0.1954 & 14.2801 $\pm$ 0.0065 & 14.2425 $\pm$ 0.0117 \\
440.6130  & 0.0092 $\pm$ 0.0002 & 0.0083 $\pm$ 0.0002 & 0.1167  & -0.3006 & 0.4172  & 13.3039 $\pm$ 0.0083 & 13.2736 $\pm$ 0.0171 \\
442.3861 & 0.0292 $\pm$ 0.0002 & 0.0281 $\pm$ 0.0003 & -0.1604 & -0.2675 & 0.1070  & 14.1004 $\pm$ 0.0040 & 14.0557 $\pm$ 0.0104 \\
442.6917 & 0.0268 $\pm$ 0.0002 & 0.0250 $\pm$ 0.0003 & -0.0209 & -0.4926 & 0.4717  & 13.9012 $\pm$ 0.0032 & 13.8699 $\pm$ 0.0080 \\
443.9546 & 0.0180 $\pm$ 0.0001 & 0.0177 $\pm$ 0.0002 & 0.8040  & 0.9715  & -0.1676 & 13.7852 $\pm$ 0.0024 & 13.7809 $\pm$ 0.0071 \\
446.3764 & 0.0197 $\pm$ 0.0001 & 0.0196 $\pm$ 0.0002 & -0.3491 & 0.2748  & -0.6239 & 13.7223 $\pm$ 0.0038 & 13.7178 $\pm$ 0.0066 \\
449.3658 & 0.0395 $\pm$ 0.0001 & 0.0393 $\pm$ 0.0002 & -1.1765 & -1.2943 & 0.1178  & 14.6131 $\pm$ 0.0116 & 14.6132 $\pm$ 0.0090 \\
450.2105 & 0.0221 $\pm$ 0.0001 & 0.0215 $\pm$ 0.0001 & 0.1513  & 0.6352  & -0.4838 & 13.8415 $\pm$ 0.0022 & 13.8373 $\pm$ 0.0050 \\
\bottomrule
	\label{Tab:Lyfeatures}
\end{longtable}

\begin{longtable}{ll ll ll ll l}
	\caption{Metal systems absorption-line parameter list -- A - complete version in electronic form}\\
    \label{metal_list_A}\\
	\toprule
	\multicolumn{1}{c}{$\lambda_{obs}$} & \multicolumn{1}{c}{Ion} & \multicolumn{1}{c}{$\lambda_{rest}$} & \multicolumn{1}{c}{z} & \multicolumn{1}{c}{$\sigma_z$} & \multicolumn{1}{c}{logN} & \multicolumn{1}{c}{$\sigma_{\text{logN}}$} & \multicolumn{1}{c}{b} & \multicolumn{1}{c}{$\sigma_{\text{b}}$} \\
	\multicolumn{1}{c}{nm} & & \multicolumn{1}{c}{nm} & & & & & \multicolumn{1}{c}{km s$^{-1}$} & \multicolumn{1}{c}{km s$^{-1}$} \\
	\midrule
	\endfirsthead
	\caption{Metal systems absorption-line parameter list - A (continued)}\\
	\toprule
	\multicolumn{1}{c}{$\lambda_{obs}$} & \multicolumn{1}{c}{Ion} & \multicolumn{1}{c}{$\lambda_{rest}$} & \multicolumn{1}{c}{z} & \multicolumn{1}{c}{$\sigma_z$} & \multicolumn{1}{c}{logN} & \multicolumn{1}{c}{$\sigma_{\text{logN}}$} & \multicolumn{1}{c}{b} & \multicolumn{1}{c}{$\sigma_{\text{b}}$} \\
	\multicolumn{1}{c}{nm} & & \multicolumn{1}{c}{nm} & & & & & \multicolumn{1}{c}{km $s^{-1}$} & \multicolumn{1}{c}{km $s^{-1}$} \\
	\midrule
	\endhead
	\midrule 
	\multicolumn{9}{r}{\footnotesize\itshape Continue on the next page}
	\endfoot
	\bottomrule
	\endlastfoot
380.3066 & O\textsc{vi}  &  103.7617  &  2.66519 &  0.00004 &  13.364 & 0.155 & 10.04 & 4.35 \\
382.1139 & Ni\textsc{ii} &  145.4842  &  1.62650 &  0.00001 &  12.991 & 0.016 & 5.17  & 0.27 \\
382.5045 & O\textsc{vi}  &  103.1926  &  2.70670 &  0.00001 &  13.401 & 0.042 & 11.46 & 1.29 \\
383.5175 & Si\textsc{ii} &  130.4370  &  1.94025 &  0.00002 &  13.594 & 0.094 & 9.27  & 2.52 \\
383.5352 & Si\textsc{ii} &  130.4370  &  1.94039 &  0.00002 &  13.019 & 0.278 & 5.29  & 1.86 \\
\end{longtable}
\begin{longtable}{ll ll ll ll l}
	\caption{Metal systems absorption-line parameter list -- B - complete version in electronic form}\\
    \label{metal_list_B}\\
	\toprule
	\multicolumn{1}{c}{$\lambda_{obs}$} & \multicolumn{1}{c}{Ion} & \multicolumn{1}{c}{$\lambda_{rest}$} & \multicolumn{1}{c}{z} & \multicolumn{1}{c}{$\sigma_z$} & \multicolumn{1}{c}{logN} & \multicolumn{1}{c}{$\sigma_{\text{logN}}$} & \multicolumn{1}{c}{b} & \multicolumn{1}{c}{$\sigma_{\text{b}}$} \\
	\multicolumn{1}{c}{nm} & & \multicolumn{1}{c}{nm} & & & & & \multicolumn{1}{c}{km s$^{-1}$} & \multicolumn{1}{c}{km s$^{-1}$} \\
	\midrule
	\endfirsthead
	\caption{Metal systems absorption-line parameter list - B (continued)}\\
	\toprule
	\multicolumn{1}{c}{$\lambda_{obs}$} & \multicolumn{1}{c}{Ion} & \multicolumn{1}{c}{$\lambda_{rest}$} & \multicolumn{1}{c}{z} & \multicolumn{1}{c}{$\sigma_z$} & \multicolumn{1}{c}{logN} & \multicolumn{1}{c}{$\sigma_{\text{logN}}$} & \multicolumn{1}{c}{b} & \multicolumn{1}{c}{$\sigma_{\text{b}}$} \\
	\multicolumn{1}{c}{nm} & & \multicolumn{1}{c}{nm} & & & & & \multicolumn{1}{c}{km $s^{-1}$} & \multicolumn{1}{c}{km $s^{-1}$} \\
	\midrule
	\endhead
	\midrule 
	\multicolumn{9}{r}{\footnotesize\itshape Continue on the next page}
	\endfoot
	\bottomrule
	\endlastfoot

385.4970 & O\textsc{vi} & 103.1926 & 2.73570 & 0.00001 & 16.639 & 0.234 & 1.82  & 0.47 \\
385.7902 & O\textsc{vi}  & 103.1926 & 2.73854 & 0.00003 & 13.592 & 0.097 & 11.87 & 3.21 \\
385.8293 & O\textsc{vi}  & 103.1926 & 2.73892 & 0.00002 & 13.809 & 0.066 & 11.43 & 1.99 \\
387.4912 & Fe\textsc{ii} & 260.0173 & 0.49025 & 0.00001 & 12.713 & 0.067 & 5.00  & 0.91 \\
387.6228 & O\textsc{vi}  & 103.7617 & 2.73570 & 0.00001 & 16.639 & 0.234 & 1.82  & 0.47 \\
\end{longtable}

\bsp	
\label{lastpage}
\end{document}